\documentclass[twocolumn,aps,superscriptaddress,nofootinbib]{revtex4-2}
\usepackage{graphicx} 
\usepackage{amsmath}
\usepackage{amssymb}
\usepackage{amsfonts}
\usepackage{bbold} 
\usepackage{enumitem}
\usepackage{braket}
\usepackage{colortbl}
\usepackage{color}
\usepackage[table]{xcolor}
\usepackage{multirow}
\usepackage[colorlinks]{hyperref}
\def\slashchar#1{\setbox0=\hbox{$#1$}
   \dimen0=\wd0 \setbox1=\hbox{/} \dimen1=\wd1
   \ifdim\dimen0>\dimen1 \rlap{\hbox to \dimen0{\hfil/\hfil}} #1
   \else  \rlap{\hbox to \dimen1{\hfil$#1$\hfil}} / \fi}
\def\p{\slashchar{p}}

\newcommand{\F}{\tilde F}
\newcommand{\G}{\tilde G}

\begin{document}
\title{Neutrino-nucleon elastic scattering in presence of non-standard
  interactions: \\cross sections and nucleon polarizations}

\author{ \surname{Ilma}} \email{ilmarafiq786@gmail.com}
\affiliation{Department of Physics, Aligarh Muslim University,
  Aligarh-202002, India}

\author{M. Rafi \surname{Alam}} \email{rafi.alam.amu@gmail.com}
\affiliation{Department of Physics, Aligarh Muslim University,
  Aligarh-202002, India}

\author{L. Alvarez-Ruso} \affiliation{Instituto de F\'isica
  Corpuscular (IFIC), Consejo Superior de Investigaciones
  Cient\'ificas (CSIC) and Universidad de Valencia, E-46980 Paterna,
  Valencia, Spain}

\author{M. Benitez \surname{Galan}} \affiliation{Departamento de F\'\i
  sica At\'omica, Molecular y Nuclear and Instituto Carlos I de F\'\i
  sica Te\'orica y Computacional, Universidad de Granada, E-18071
  Granada, Spain}

\author{I. Ruiz \surname{Simo}} \email{ruizsig@ugr.es}
\affiliation{Departamento de F\'\i sica At\'omica, Molecular y Nuclear
  and Instituto Carlos I de F\'\i sica Te\'orica y Computacional,
  Universidad de Granada, E-18071 Granada, Spain}

\author{S. K. \surname{Singh}} \affiliation{Department of Physics,
  Aligarh Muslim University, Aligarh-202002, India}

\begin{abstract} 
New physics beyond the Standard Model (SM) may appear in the form of
non-standard neutrino interactions (NSI). We have studied neutral
current (anti)neutrino-nucleon scattering in presence of NSI. We
obtain that in this scenario, nucleon matrix elements depend not only
on the isovector axial nucleon form factor but also on the isoscalar
one. For the axial form factors we consequently rely on the quark
flavor decomposition performed by QCD simulations in the lattice
(LQCD).  We have examined cross sections and polarization
observables. For the current bounds on diagonal muon flavor NSI
couplings we find substantial deviations from the SM predictions in
cross sections and transverse polarizations of the outgoing
nucleons. In view of the progress in the precision of LQCD
determinations of nucleon properties, modern measurements of neutral
current (anti)neutrino-nucleon scattering will be in the position to
discover or significantly constrain NSI.
\end{abstract}
\maketitle

\section{Introduction}

The Standard Model (SM) is a successful paradigm for understanding
particle physics.  In particular, it has proved capable of explaining
a wide range of physical phenomena involving electroweak
interactions. The SM assumes neutrinos of all flavors to be massless
and the conservation of lepton flavors in neutrino
interactions~\cite{Weinberg:1967tq}. However, it has been established
that neutrinos undergo flavor oscillations during their propagation,
thus violating the lepton flavor conservation~\cite{Workman:2022ynf}.
The phenomenon of neutrino oscillation further implies that neutrinos
are not massless, and the mass states are nontrivial admixtures of
flavor states.  The theoretical models explaining this physics go
beyond the SM and, below the weak scale, give rise to new types of
neutrino
interactions~\cite{Lindner:2016wff,Bischer:2019ttk,Han:2020pff,Du:2021rdg}. The
subset of these generalized interactions involving only SM
left(right)-handed (anti)neutrinos and Lorentz-invariant four-fermion
terms built from vector and axial-vector operators is
phenomenologically coined as non-standard interactions
(NSI)~\cite{Wolfenstein:1977ue,Grossman:1995wx,Berezhiani:2001rs,Davidson:2003ha,Ohlsson:2012kf}.

Non-standard interactions are also categorized into two groups,
Charged Current (CC) NSI and Neutral Current (NC) NSI, similar to
standard weak interactions.  Through matter effects, certain
combinations of NC NSI can bias some of the mixing angles extracted
from oscillation experiments (for a review see
Ref.~\cite{Farzan:2017xzy}). Moreover, besides measuring mixing
parameters with unprecedented accuracy, the upcoming generation of
neutrino experiments will be sensitive to yet-unknown neutrino
parameters like the mass ordering and CP-violating phase. NSI can have
an impact on the interpretations of long baseline experiments
regarding the CP violating
phase~\cite{Masud:2015xva,deGouvea:2015ndi,Forero:2016cmb,Esteban:2019lfo,Capozzi:2019iqn}. Therefore,
the study of NSI is crucial for accurately interpreting the results of
neutrino experiments and understanding the fundamental properties of
neutrinos.

Non-standard interactions can also modify the way neutrinos scatter on
matter, potentially leading to deviations from the SM predictions. The
NSI contribution to neutrino-electron cross sections has been studied
and constrained using experimental
data~\cite{Barranco:2007ej,Forero:2011zz,Khan:2014zwa,Escrihuela:2021mud}. Upper
limits for the NSI parameters in the quark sector have been obtained
from neutrino-nucleus scattering data in the deep inelastic scattering
regime probed at high
energies~\cite{Davidson:2003ha,Escrihuela:2011cf,Escrihuela:2021mud}. Further
constraints on vector NC NSI couplings can be extracted from
measurements for neutrino-nucleus elastic scattering, as has already
been done with COHERENT
data~\cite{COHERENT:2017ipa,Coloma:2017ncl,Liao:2017uzy,AristizabalSierra:2018eqm,Denton:2020hop,Khan:2021wzy,Flores:2021kzl,DeRomeri:2022twg}. The
impact of NSI on neutrino-nucleon (quasi)elastic and inelastic cross
sections and related observables is considerably less
explored~\cite{Papoulias:2016edm,Borah:2024hvo,Tomalak:2024yvq}. This
is justified by the present uncertainties in the hadronic input and
the lack of precise data which render the task of extracting
information about NSI highly difficult. Nevertheless, the availability
of new Lattice QCD (LQCD) determinations of the isovector nucleon
axial form factor (see Ref.~\cite{Meyer:2022mix} and references
therein) together with the prospects of modern experimental studies of
neutrino-nucleon
interactions~\cite{Alvarez-Ruso:2022ctb,Duyang:2024ucj} following the
novel extraction of $\bar\nu_\mu \, p \rightarrow \mu^+ \, n$-proton
cross section by the MINERvA experiment~\cite{MINERvA:2023avz} provide
motivations to explore this avenue. In fact, a recent analysis of
these data obtained new confidence intervals for tensor and scalar
interactions~\cite{Tomalak:2024yvq}. Furthermore,
Ref.~\cite{Kopp:2024yvh} finds that tensor and pseudoscalar
interactions might enhance neutrino-nucleus CC quasielastic scattering
cross sections on nuclei.  Along these lines, the $n-\Lambda_c$
transition induced by $\nu_\tau$ CC interactions in the presence of
new physics has been recently examined~\cite{Kong:2023kkd}, showing
how results from LQCD help to explore the new physics backed by the
future experimental data.

In this work, we investigate the impact of NSI
on(anti)neutrino-nucleon NC elastic (NCE) cross sections, improving
and extending the study of Ref.~\cite{Papoulias:2016edm}, and explore
their effects in spin-dependent observables for the first time in the
NC case. In particular, we study the polarization of the outgoing
nucleon in the presence of NSI. There is no doubt that polarization
measurements are challenging, even more so in the case of neutrino
interactions, but spin observables allow for a more detailed scrutiny
of the dynamics in play, as shown in several theoretical studies. The
polarization of the outgoing lepton in CC quasielastic and inelastic
neutrino-nucleon and neutrino-nucleus scattering has been extensively
investigated~\cite{Hagiwara:2003di,Graczyk:2004uy,Kuzmin:2004yb,Kuzmin:2004ya,Valverde:2006yi,Sobczyk:2019urm,Fatima:2020pvv,Hernandez:2022nmp,Isaacson:2023gwp}. This
polarization is relevant for the $\nu_\tau$ appearance
measurements~\cite{Li:2017dbe}. The asymmetries arising from the
polarization of target nucleon and outgoing nucleon and lepton are
derived in Ref.~\cite{Borah:2024hvo} for CC quasielastic scattering in
presence of generalized interactions. The polarization of hyperons
produced in CC neutrino-nucleon scattering has also been
studied~\cite{Akbar:2016awk,Fatima:2018tzs}. It has been found that
$T$-violating interactions lead to hyperon polarizations perpendicular
to the scattering plane.  For weak pion production, the sensitivity of
polarization asymmetries to details of the interaction, such as the
relative phases between resonant and non-resonant amplitudes, has been
addressed in Refs.~\cite{Graczyk:2021oyl,Graczyk:2023lrm}.

For our study, we adopt the formalism of
Refs~\cite{Bilenky:2013fra,Bilenky:2013iua}, where the final-nucleon
polarization in NCE and CC quasielastic neutrino-nucleon scattering
was calculated. The critical difference resides on the fact that the
motivation then was to obtain information about the axial and strange
axial form factors of the nucleon from neutrino scattering
observables. Instead, we assume that the required nonperturbative QCD
input can be (now or in the near future) provided by LQCD with
sufficient precision to render feasible the investigation of new
physics in neutrino-nucleon elastic scattering with future
experimental data.  From this perspective, the present paper can be
regarded as a generalization of Ref.~\cite{Bilenky:2013iua} to the NSI
scenario. It is organized as follows: in Sec.~\ref{sec:formalism} the
formalism is presented. We introduce NC neutrino interactions with
quarks in presence of NSI, which are then used to derive the nucleon
matrix elements. Our choice for the nucleon form factors and NSI
couplings are given afterwards. Next we introduce the spin density
matrix, the cross section and derive the polarization of the outgoing
nucleon. Our numerical results are presented in
Sec.~\ref{sec:results}. We conclude afterwards in
Sec.~\ref{sec:conclusions}.

\section{Formalism}\label{sec:formalism}

\subsection{Neutrino-quark NC interactions in the presence of NSI}

Following the common notation in the literature (see for instance
Ref. \cite{Miranda:2015dra}), the effective four-fermion NC-NSI
Lagrangian is given by
\begin{equation}
    \label{eq:NC-NSI}
    \mathcal{L}^\mathrm{NC}_{\mathrm{NSI}} = - 2 \sqrt{2} G_F
    \epsilon^{f X}_{ij} \bar{\nu}^i \gamma_\mu P_L \nu^j \bar{f}
    \gamma^\mu P_X f \,,
\end{equation}
in terms of the projectors $P_{X=L,R} = (1 \mp \gamma_5)/2$; ${i,j} =
e\,,\mu\,,\tau$ are neutrino flavor indices; $G_F$ is the Fermi
constant. In general $f$ denotes lepton or quark fields. The present
study is restricted to neutrino interactions with light quarks: $f=
u,d, s$. Quark flavor conservation is implicit in the
notation. Instead, the NC NSI couplings, denoted $\epsilon^{f
  X}_{ij}$, can be flavor diagonal, $i=j$, or flavor changing, $i \neq
j$, for the neutrinos.

In the limit where both NSI and SM NC interactions can be treated as
contact ones, it is convenient to write an effective Lagrangian
encompassing both SM and NSI
\begin{equation}
    \label{eq:NC-SM+NSI}
    \mathcal{L}^\mathrm{NC}_{\mathrm{SM+NSI}} = - \frac{G_F}{\sqrt{2}}
    \, L_\mu^{ij} J^\mu_{ij} \,
\end{equation}
with
\begin{equation}
    \label{eq:leptonNC}
    L_\mu^{ij} = \bar{\nu}^i \gamma_\mu (1-\gamma_5) \nu^j \,,
\end{equation}
and
\begin{align}
    \label{eq:quarkNC}
    J^\mu_{ij}&=
    \delta_{ij}\;
    J^\mu_{\rm SM} +
    \left(J^\mu_{ij}\right)_{\rm NSI}\,,
    \nonumber\\
    J^\mu_{\rm SM}&=
    \bar{u}\gamma^\mu
    \left[
    \frac12-\left(\frac23
    \right)2s^2_w-\frac12
    \gamma_5
    \right]u
    \nonumber\\
    &+\bar{d}\gamma^\mu 
    \left[
    -\frac12-\left(
    -\frac13\right)2s^2_w+
    \frac12 \gamma_5
    \right]d
    \nonumber\\
    &+\bar{s}\gamma^\mu 
    \left[
    -\frac12-\left(
    -\frac13\right)2s^2_w+
    \frac12 \gamma_5
    \right]s \,,
    \nonumber\\
    \left(J^\mu_{ij}
    \right)_{\rm NSI}&=
    \bar{u}\gamma^\mu\left(
    \epsilon^{uV}_{ij}-
    \epsilon^{uA}_{ij}
    \gamma_5
    \right)u \nonumber\\
    &+ \bar{d}\gamma^\mu \left(
    \epsilon^{dV}_{ij}-
    \epsilon^{dA}_{ij}
    \gamma_5
    \right)d \nonumber\\
    &+ \bar{s}\gamma^\mu \left(
    \epsilon^{sV}_{ij}-
    \epsilon^{sA}_{ij}
    \gamma_5
    \right)s 
    \,,
\end{align}
where we have explicitly split the contribution of the quarks NC into
the SM lepton-flavor diagonal current, and the, in general, non flavor
diagonal NSI NC. In eqs.~(\ref{eq:quarkNC}) $u,d$ and $s$ denote the
spinor fields of the corresponding quark flavors; $\epsilon^{V,A} =
\epsilon^L \pm \epsilon^R$ and $s_w \equiv \sin{\theta_w}$, where
$\theta_w$ is the weak mixing angle.

\subsection{Neutrino-nucleon NC interactions in presence of NSI}

The (anti-)neutrino induced NCE scattering processes are: 
    \begin{align}\label{eq:reac}
        \overset{(-)}{\nu} (k,0) + N (p, M) &\rightarrow
        \overset{(-)}{\nu} (k^\prime, 0) + N (p^\prime, M),
    \end{align}
where $N(\equiv p,n)$ stands for the nucleon, and the corresponding
four momenta with their masses are given in parentheses. The
scattering amplitude is given by,
\begin{equation}
    \label{eq:Ampl}
    \mathcal{M} = \frac{G_F}{\sqrt{2}} l_\mu^{ij} \braket{ N |
      J^\mu_{ij} | N } \,
\end{equation}
where 
\begin{equation}
    \label{eq:nucurrent}
    l_\mu^{ij} = \bar u^i(k') \gamma_\mu (1 - \gamma_5 ) u^j(k)
\end{equation}
or 
\begin{equation}
    \label{eq:anucurrent}
    l_\mu^{ij} = \bar v^j(k) \gamma_\mu (1 - \gamma_5 ) v^i(k')
\end{equation}
for neutrinos or antineutrinos, respectively; $u,v(k)$ denote spinors
in momentum space. Assuming isospin symmetry ($m_u=m_d \neq m_s$) and
neglecting radiative corrections\footnote{QED radiative corrections
for CC neutrino-nucleon quasielastic scattering have been recently
calculated in Refs.~\cite{Tomalak:2021hec,Tomalak:2022xup}.}, the
nucleon matrix element
\begin{equation}
    \label{eq:nucleonMatEl}
    \braket{ N | J^\mu_{ij} | N } = \bar{u}(p')  \Gamma^\mu_{ij} u(p)
\end{equation}
can be expressed in terms of vector and axial-vector form factors
(FF):
\begin{align}
        \Gamma^\mu &= V^\mu-A^\mu\; , \nonumber \\ V^\mu &\equiv
        \tilde{F}^{(N)}_1 (Q^2)\gamma^{\mu} + i
        \frac{\tilde{F}^{(N)}_2(Q^2)}{2M}\sigma^{\mu \nu} q_{\nu} \; ,
        \nonumber \\ A^\mu &\equiv \tilde{F}^{(N)}_A(Q^2) \gamma^{\mu}
        \gamma_5 + \frac{\tilde{F}^{(N)}_P(Q^2)}{M} q^{\mu} \gamma_5
        \; ,
        \label{eq:had_vert}
\end{align}
which depend upon the four-momentum transfer squared
$Q^2=-(k-k^\prime)^2$. Lepton flavor indices have been dropped to
alleviate the notation. In the observable quantities, the contribution
of $\tilde{F}_P$ is always proportional to powers of the outgoing
neutrino mass and, therefore, negligible. It is actually exactly zero
in the massless neutrino approximation adopted here and disregarded
from now on. Owing to isospin symmetry, the FF introduced in
Eq.~(\ref{eq:had_vert}) can be related to the isovector and isoscalar
vector and axial vector FF. Furthermore, the vector ones can be
written in terms of the electromagnetic FF of protons and neutrons
$F_{1,2}^{(p,n)}$. In the presence of NSI, one obtains
that\footnote{Although with a different notation, this equation
coincides with Eqs.~(24-27) of the preprint~\cite{Abbaslu:2024jzo},
which appeared in the arXiv a few days after ours.}
\begin{align}
    \label{eq:NCFFNSI}
     \tilde{F}^{(p,n)}_{1,2} &= \left[ \left(\frac{1}{2} - 2
       s_w^2\right)\delta_{ij} + 2 \epsilon^{uV}_{ij} +
       \epsilon^{dV}_{ij} \right] F_{1,2}^{(p,n)} \nonumber \\ &+
     \left( - \frac{1}{2}\delta_{ij} + 2 \epsilon^{dV}_{ij} +
     \epsilon^{uV}_{ij} \right) F_{1,2}^{(n,p)} \nonumber \\ &+ \left(
     - \frac{1}{2} \delta_{ij} + \epsilon^{uV}_{ij} +
     \epsilon^{dV}_{ij} + \epsilon^{sV}_{ij} \right) F_{1,2}^s \,,
     \nonumber \\ \tilde{F}^{(p,n)}_A &= \pm \frac{1}{2} \left(
     \delta_{ij} + \epsilon^{uA}_{ij} - \epsilon^{dA}_{ij} \right)
     F^{iv}_A \nonumber \\ &+ \frac{1}{2} \left( \epsilon^{uA}_{ij} +
     \epsilon^{dA}_{ij} \right) F^{is}_A \nonumber \\ &+ \left( -
     \frac{1}{2} \delta_{ij} + \epsilon^{sA}_{ij} \right) F_A^s \,.
\end{align}
In this way, the NSI contributions to the (anti)neutrino-nucleon cross
sections are shaped by the well defined nucleon FF, which can be
obtained from experiment and/or LQCD.\footnote{There is no need to
introduce ad-hoc $Q^2$ dependent FF as in
Ref.~\cite{Papoulias:2016edm}.}

Remarkably, in presence of NSI, the nucleon NC matrix element depends
not only on the standard axial isovector FF, $F^{iv}_A$:
\begin{align}
    \label{eq:isoVaxial}
    &\braket{ p,n | \left( \bar{u} \gamma^\mu \gamma_5 u - \bar{d}
      \gamma^\mu \gamma_5 d \right)| p,n } \equiv \nonumber \\ &\pm
    \bar{u}(p') \left[ F^{iv}_A(Q^2) \gamma^{\mu} \gamma_5 +
      \frac{F^{iv}_P(Q^2)}{M} q^{\mu} \gamma_5 \right] u(p)
\end{align}
but also on the axial vector isoscalar FF, $F^{is}_A$:
\begin{align}
    \label{eq:isoSaxial}
    &\braket{ p,n | \left( \bar{u} \gamma^\mu \gamma_5 u + \bar{d}
      \gamma^\mu \gamma_5 d \right)| p,n } \equiv \nonumber
    \\ &\bar{u}(p') \left[ F^{is}_A(Q^2) \gamma^{\mu} \gamma_5 +
      \frac{F^{is}_P(Q^2)}{M} q^{\mu} \gamma_5 \right] u(p),
\end{align}
not probed by SM electroweak interactions.

\subsection{Form factors}

Assuming there is no sizable new physics in electron scattering, one
can take the electromagnetic FFs $F_{1,2}^{(p,n)}$ determined by the
electron scattering and parameterized using experimental data. In
fact, here we adopt standard parametrizations presented in
Appendix~\ref{app:ff}.  The non-zero contribution due to the strange
quark-sea in the nucleon leads to the strange vector ($F_{1,2}^s$) and
axial ($F_A^s$) FF~\cite{Maas:2017snj}. For the strange vector FF, we
adopt the parametrization given in Ref.~\cite{Garvey:1992cg},
\begin{align}
    F_1^s (Q^2) &= -\frac{1}{6} \langle r_s^2 \rangle Q^2 F(Q^2)
    \nonumber \\ F_2^s (Q^2) &= \mu_s F(Q^2) \, ,
\end{align}
where $\langle r_s^2 \rangle$ is the mean-squared strange radius of
the nucleon and $\mu_s$ is the strange magnetic moment. The $\langle
r_s^2 \rangle$ can be further expressed in terms of the mean-squared
strange charge radius, $\langle r_E^2 \rangle^s$, and $\mu_s$ as
\begin{align}
    \langle r_s^2 \rangle &= \langle r_E^2 \rangle^s
    -\frac{3}{2}\frac{\mu_s}{M^2} \,.
\end{align}
A modified dipole form factor
\begin{equation} 
    F(Q^2) = \left( 1+\frac{Q^2}{4M^2} \right)^{-1} \left(
    1+\frac{Q^2}{m_v^2} \right)^{-2}
\end{equation}
is taken for the $Q^2$ dependence. The dipole mass, $m_v$, can be
related to $\langle r_E^2 \rangle^s$ and the strange magnetic radius
$\langle r_M^2 \rangle^s$ as\footnote{The feature of dipole
parametrizations that the whole $Q^2$ dependence is constrained by the
low-$Q^2$ behavior is not justified by QCD. Nevertheless, given the
insufficient knowledge available about the strange FF of the nucleon,
the adopted dipole parametrization suffices for the present
exploratory study.}
\begin{equation}
    m_v^2 = 12 \mu_s (\langle r_m^2 \rangle^s - \langle r_E^2
    \rangle^s )^{-1} \, .
\end{equation} 

A priori, no conclusive bounds are available from experiments for any
of these parameters~\cite{Maas:2017snj}. On the other hand,
significant progress has been made in the determination of nucleon
properties on the lattice in the past few years (see for instance the
FLAG review~\cite{flavourLatticeAveragingGroupFLAG:2021npn}). We adopt
the numerical values of these parameters obtained in
Ref~\cite{Alexandrou:2019olr} by the Extended Twisted Mass (ETM)
Collaboration: $\mu_s = -0.017(4)$, $\langle r_M^2 \rangle^s
=-0.015(9)$ fm$^2$ and $\langle r_E^2 \rangle^s =-0.0048(6)$ fm$^2$.

Because of its interest in neutrino physics, the isovector axial FF
has received considerable attention from LQCD practitioners, and
several determinations have become
available~\cite{Meyer:2022mix}. However, as shown in
Eq.~(\ref{eq:NCFFNSI}), in presence of NSI more input from QCD is
required. The flavor decomposition of the axial current
\begin{align}
    \label{eq:flavoraxial}
    &\braket{ N | \bar{q} \gamma^\mu \gamma_5 q| N } \equiv \nonumber
    \\ &\bar{u}(p') \left[ F^{q}_A(Q^2) \gamma^{\mu} \gamma_5 +
      \frac{F^{q}_P(Q^2)}{M} q^{\mu} \gamma_5 \right] u(p) \,,
\end{align}
with $q=u,d,s$, performed also by the ETM
collaboration~\cite{Alexandrou:2021wzv} is well suited for our
purposes. They provide dipole parametrizations
\begin{align}
    F_A^{q}(Q^2) = g_A^{q} \left( 1+ \frac{Q^2}{m_{A q}^{2}}
    \right)^{-2} \,.
\end{align}
In Ref.~\cite{Alexandrou:2021wzv}, more realistic z-expansion fits
were also performed, although their parameters are not disclosed in
the publication. Furthermore, the results of the z-expansion fits
largely overlap with the dipole ones for $u$ and $d$ quarks, as can be
seen in Fig. 15 of Ref.~\cite{Alexandrou:2021wzv}. For these reasons,
we rely on the dipole parametrizations although the search for NSI in
(future) data will benefit from the best possible input from LQCD.

Couplings $g_A^q$ and axial masses $m_{A q}^2$ from
Ref.~\cite{Alexandrou:2021wzv} are compiled in
Table~\ref{tab:axialflav_para}.
\begin{table}[t]
  \centering
  \renewcommand{\arraystretch}{1.25}
       \begin{tabular}{|c|c|c|} \hline
        $q$ & $g_A^q$ & $m_{A q}$ [GeV] \\ \hline \hline
         $u$ & 0.859(18) & 1.187(65)  \\ 
         $d$ &  -0.423(17) & 1.168(54) \\
         $s$ & -0.044(8) & 0.992(164)  \\ \hline       
        \end{tabular}
   \caption{Parameters of the flavor decomposition of the nucleon
     axial form factor according to Ref.~\cite{Alexandrou:2021wzv}.}
   \label{tab:axialflav_para}
\end{table}
It is then straightforward to find 
\begin{align}
    F_A^{iv}(Q^2) &= F_A^u(Q^2) - F_A^d(Q^2) \nonumber \\
    F_A^{is}(Q^2) &= F_A^u(Q^2) + F_A^d(Q^2) \, .
\end{align}  

 The errors in the parameters introduced in this section are used to
 assess the impact of LQCD uncertainties on the observables. We
 consider these errors to be uncorrelated. Therefore, the resulting
 bands should be taken with caution as they do not incorporate the
 correlations that are likely to exist among the LQCD FF.

\subsection{NSI couplings}

In the present study, NSI couplings are assumed to be real,
disregarding CP (or T) violation~\cite{Cabibbo:1964zza}. Furthermore,
we restrict ourselves to the flavor diagonal case and focus on muon
(anti)neutrinos, which represent the dominant fraction in accelerator
beams. Finally, we can safely neglect NSI couplings to strange quarks,
for which there are no bounds at present. Besides being small, these
$\epsilon^{sX}_{ij}$ appear always multiplied by the strange FF of the
nucleon which are also quite small.

All in all, we are left with $\epsilon^{uV,dV}_{\mu\mu}$ and
$\epsilon^{uA,dA}_{\mu\mu}$; 90\% confidence level (CL) intervals have
been obtained from the analysis of atmospheric neutrino oscillations
at Super-Kamiokande and high energy scattering data taken at
NuTeV~\cite{Escrihuela:2011cf} and are given in
Table~\ref{tab:nsi_para}. In the absence of information about the
correlation matrices, we treat them as uncorrelated.

\begin{table}[h!]
   \centering 
   \renewcommand{\arraystretch}{1.25}
       \begin{tabular}{|c|c|c|c|} \hline
        NSI & 90\% CL & NSI & 90\% CL\\ Couplings & interval &
        Couplings & interval \\ \hline \hline $\epsilon^{uV}_{\mu
          \mu}$ & [--0.044,0.044] & $\epsilon^{uA}_{\mu \mu}$ &
        [--0.094,0.14] \\ $\epsilon^{dV}_{\mu \mu}$ & [--0.042,0.042]
        & $\epsilon^{dA}_{\mu \mu}$ & [--0.072,0.057] \\ \hline
        \end{tabular}
   \caption{Neutral current flavor diagonal vector and axial
     neutrino-quark NSI couplings taken from
     Refs.~\cite{Escrihuela:2011cf,Farzan:2017xzy}.}
   \label{tab:nsi_para}
\end{table}

\subsection{Spin Density Matrix}

After the neutrino-nucleon interaction, the polarization state of the
outgoing nucleon is characterized by its spin density
matrix~\cite{Bilenky:1995zq,Bilenky:2013fra,Bilenky:2013iua}:
    \begin{align}
        \rho_f(p^\prime) = \mathcal{L}^{\mu\nu} \Lambda(p^\prime)
        \Gamma_\mu \, \rho(p) \, \Tilde{\Gamma}_\nu \Lambda(p^\prime)
        \,,
    \end{align}
    where $\tilde{\Gamma}_\nu=\gamma^0 \Gamma_\nu^\dagger \gamma^0$,
    with $\Gamma_\nu$ introduced in Eq. (\ref{eq:had_vert}), and
    $\Lambda(p) \equiv \p + M$.  The leptonic tensor, obtained from
    the currents in Eqs. (\ref{eq:nucurrent},\ref{eq:anucurrent}) is
    \begin{align}
        \mathcal{L}^{\mu \nu} = {\rm{\textbf{Tr}}}[\gamma^\mu (1\mp
          \gamma_5) \slashchar{k} \gamma^\nu (1\mp \gamma_5)
          \slashchar{k}'],
    \end{align}
    where the upper (lower) sign is for neutrinos
    (anti-neutrinos). Assuming that the initial nucleon is unpolarized
    \begin{align}
        \rho(p) = \frac12 \Lambda(p)
    \end{align}

  \subsection{Differential cross section}
    The differential cross section for the process in (\ref{eq:reac})
    is given by
    \begin{align}\label{eq:diffxsec}
        \frac{d \sigma}{dQ^2}=\frac{1}{64\pi M^2 E_\nu^2}  
        \overline{ \sum} \sum |\mathcal M|^{2}
    \end{align}
    where $E_\nu$ is the incoming (anti)neutrino energy in the
    Laboratory frame (LAB). The scattering amplitude of
    Eq.~(\ref{eq:Ampl}) squared is summed over the final nucleon spins
    and averaged over the initial ones:
    \begin{align}
        \overline{\sum} \sum |{\cal{M}}|^2 = \frac{G_F^2}{4}
        \mathcal{L}^{\mu \nu} \mathcal{J}_{\mu \nu}
    \end{align}  
    
    The hadronic tensor $\mathcal{J}_{\mu \nu}$ is written with the
    help of Eqs.~(\ref{eq:nucleonMatEl},\ref{eq:had_vert}) as,
    \begin{align}\label{eq:J}
        \mathcal{J}_{\mu \nu}&={\rm{\bf Tr}} \left[ \Gamma_\mu
          \Lambda(p) \tilde{\Gamma}_\nu \Lambda(p^\prime) \right]
        \,. \end{align}
    Using the identity
\begin{align}
\label{eq:proy}
\Lambda(p^\prime)\Lambda(p^\prime) = 2 M \Lambda(p^\prime)\,, 
\end{align}
$\mathcal{J}_{\mu \nu}$ can be recast as
\begin{align}
        \label{eq:LJ}
        \mathcal{J}_{\mu \nu}&= \frac{1}{2M} {\rm{\bf Tr}} \left[
          \Lambda(p^\prime) \Gamma_\mu \Lambda(p) \tilde{\Gamma}_\nu
          \Lambda(p^\prime) \right] \nonumber \\ &= \frac{1}{M}
                {\rm{\bf Tr}} \left[ \Lambda(p^\prime) \Gamma_\mu
                  \rho(p) \tilde{\Gamma}_\nu \Lambda(p^\prime) \right]
                \,.
        \end{align}
Therefore, 
\begin{align}
\label{eq:N}
 \mathcal{L}^{\mu \nu} \mathcal{J}_{\mu \nu} &= \frac{1}{M} {\rm{\bf
     Tr}} \left[ \rho_f(p^\prime) \right] \equiv 64 M^2 E_\nu^2
 \mathcal{N} \,.
    \end{align}  

\subsection{Final-nucleon polarization}
    The polarization 4-vector for the final nucleon corresponding to
    reaction (\ref{eq:reac}) can be written in terms of the spin
    density matrix
    as~\cite{Bilenky:1995zq,Bilenky:2013fra,Bilenky:2013iua}:
    \begin{align}\label{eq:pol1}
        \zeta^\tau = \frac{{\rm{\bf Tr}} [\gamma^\tau \gamma_5
            \rho_f(p^\prime)]}{{\rm{\bf Tr}} [\rho_f(p^\prime)]} =
        \frac{\mathcal{L}^{\alpha \beta} {\rm{\bf Tr}}[\gamma^\tau
            \gamma_5 \Lambda(p^\prime) \Gamma_\alpha \Lambda(p)
            \tilde{\Gamma}_\beta \Lambda(p^\prime) ]}{
          \mathcal{L}^{\alpha \beta} {\rm{\bf Tr}} [\Lambda(p^\prime)
            \Gamma_\alpha \Lambda(p) \tilde{\Gamma}_\beta
            \Lambda(p^\prime) ]}
    \end{align}
    We then rely on the identities, 
    \begin{align}
        \Lambda(p^\prime) \gamma^\tau \gamma_5 \Lambda(p^\prime) &= 2
        M \left( g^{\tau \sigma} - \frac{p^{\prime \tau} p^{\prime
            \sigma}}{M^2} \right) \Lambda(p^\prime) \gamma_\sigma
        \gamma_5 \,,
    \end{align} 
and Eq. (\ref{eq:proy}) to rewrite Eq.~(\ref{eq:pol1}) as
\cite{Bilenky:2013fra,Bilenky:2013iua}
    \begin{align}\label{eq:pol2}
        \zeta^\tau = \left( g^{\tau \sigma} -
        \frac{p^{\prime\tau}p^{\prime \sigma}}{M^2} \right)
        \frac{\mathcal{L}^{\alpha\beta} {\rm{\bf Tr}}[\gamma_\sigma
            \gamma_5 \Lambda(p^\prime) \Gamma_\alpha
            \Lambda(p)\Tilde{\Gamma}_\beta]}{\mathcal{L}^{\alpha\beta}
          {\rm{\bf Tr}}[\Lambda(p^\prime) \Gamma_\alpha
            \Lambda(p)\Tilde{\Gamma}_\beta]} \, .
    \end{align}
 Using Eq.~(\ref{eq:pol2}), it is easy to verify that in the rest
 frame of the final nucleon the polarization vector is purely
 space-like, i.e. $\zeta^\tau = (0, \vec \zeta)$. Further,
 Eq.~(\ref{eq:pol2}) explicitly shows that $\zeta \cdot p^\prime = 0$.

\begin{figure*}
    \centering
    \includegraphics[scale=0.99]{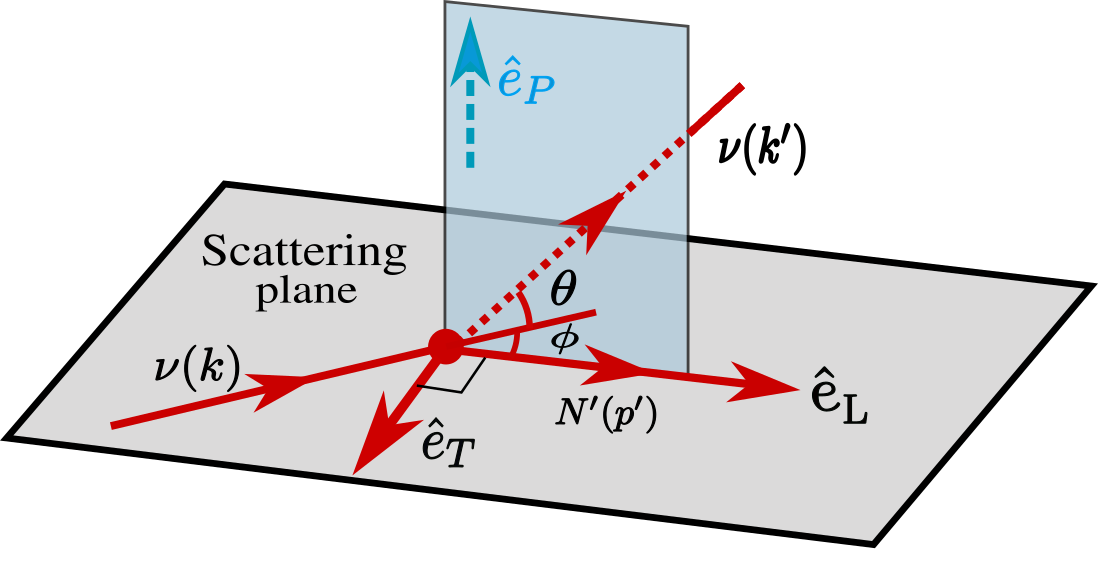}
    \caption{Kinematics for the NCE process in the laboratory frame,
      where $\hat e_L$, $\hat e_P$ and $\hat e_T$ represent
      longitudinal, perpendicular and transverse directions of the
      final nucleon polarization vector. Indicated are also the
      directions of the incoming neutrino and the outgoing neutrino
      and nucleon, with their corresponding scattering angles $\theta$
      and $\phi$.}
    \label{fig:pol}
\end{figure*}

In the following, we first obtain the polarization 4-vector in LAB,
where it is denoted as $\chi$, by evaluating Eq.~(\ref{eq:pol2}) in
that frame. Then, it is expressed in the rest frame of the final
nucleon by means of the boost
\begin{align}\label{eq:pol3}
    \zeta^\sigma =
    \Lambda_\tau^\sigma(\vec{\beta}=\vec{p}^{\,\prime}/E_{p'}) \,
    \chi^\tau ,
\end{align}
where $E_{p'} = \sqrt{M^2 + \vec{p}^{\,\prime \,2}}$. The kinematics
for the reaction in LAB is depicted in Fig.~\ref{fig:pol}. As the
target is unpolarized and the incident particles are polarized along
the direction of their momentum, the azimuthal dependence averages
out. Therefore, all momenta lie on the scattering plane.

We write the polarization vector ($\vec \chi$)\footnote{The time
component $\chi^0$ can be readily obtained from the orthogonality
condition $\chi\cdot p'=0$.} in terms of laboratory variables,
spanning it along three basis vectors defined as:
\begin{align}\label{eq:pol_basis}
    \hat e_L &= \frac{\vec{p}^{\;\prime}}{|\vec{p}^{\;\prime}|}
    \nonumber \\ \hat e_P &= \frac{\vec{p}^{\;\prime} \times
      \vec{k}}{| \vec{p}^{\;\prime} \times \vec{k} | } \nonumber
    \\ \hat e_T &= \hat e_L \times \hat e_P
\end{align}
In such a basis, 
\begin{align}\label{eq:chi}
    \vec{\chi} = \chi_L \hat{e}_L + \chi_T \hat{e}_T + \chi_P
    \hat{e}_P \, ,
\end{align}
where the components along longitudinal, perpendicular and transverse
directions are denoted $\chi_L$, $\chi_P$ and $\chi_T$, respectively.

Figure~\ref{fig:pol}, illustrates that, while the longitudinal and
transverse components lie on the scattering plane, the perpendicular
component is orthogonal to the scattering plane.  The perpendicular
component arises in presence of $T$
violation~\cite{Fatima:2018gjy,SajjadAthar:2022pjt}. In the present
study the NSI couplings are taken to be real, so $T$ invariance is
respected and $\chi_P = 0$.

In order to extract the components $\chi_{L,T}$ we first write the
most general expression for $\vec \chi$ using the available vectors:
\begin{align}\label{eq:nxi}
  \mathcal{N}  \vec{\chi} = \tilde a_k \vec{k} + \tilde a_{p^\prime} \vec{p}^{\;\prime} + \tilde a_{k^\prime} \vec{k}^\prime + \tilde a_p \vec{p} \,, 
\end{align}
where coefficients $\tilde a_i$ are Lorentz scalars. In the LAB frame
$\vec{p}=0$ and, hence, the contribution due to $\tilde a_p$ is
zero. Finally, momentum conservation reduces the number of independent
terms by one.  Therefore we redefine Eq.~(\ref{eq:nxi}) as
\begin{align}\label{eq:chi2}
  \mathcal{N} \vec{\chi} = a_k \vec{k} + a_{p^\prime}
  \vec{p}^{\;\prime} \,.
\end{align}
Using Eqs.~(\ref{eq:pol_basis})--(\ref{eq:chi2}), 
one can easily obtain 
    \begin{align}\label{eq:chi_3}
        \chi_L(Q^2) & = \frac{[a_k \vec{k} \cdot \vec{p}^{\;\prime} +
            a_{p^\prime} |\vec{p}^{\;\prime}|^2]}{|\vec{p}^{\;\prime}|
          \mathcal{N}} \nonumber \\ \chi_T(Q^2) &= \frac{a_k[(\vec{k}
            \cdot \vec{p}^{\;\prime})^2 - |\vec{k}|^2
            |\vec{p}^{\;\prime}|^2
        ]}{\mathcal{N}|\vec{p}^{\;\prime}||\vec{k} \times
          \vec{p}^{\;\prime}|}\,.
    \end{align}

    As announced earlier, we use the boost of Eq.~(\ref{eq:pol3}) to
    get the polarization vector $\vec \zeta$ in the final nucleon's
    rest frame. As the Lorentz boost is along the direction of $\vec
    p^{\;\prime}$ or $\hat e_L$, the transverse component being
    orthogonal to $\hat e_L$ will not be affected, i.e.  $\zeta_T =
    \chi_T$. In the longitudinal direction, we have
\begin{align*}
   {\chi}_L =\gamma ( {\zeta}_L - \beta \zeta_0)
\end{align*}
where $\gamma = E_{p^\prime}/M$ and $\zeta_0 = 0$. Therefore the final
expressions for polarization components in the rest frame of the final
nucleon, given in the basis of Eq. (\ref{eq:pol_basis}) read as:
\begin{align}\label{eq:long/transpol}
    \zeta_L(Q^2) & = \frac{M}{E_{p^\prime}} [a_k \vec{k} \cdot
      \vec{p}^{\;\prime} + a_{p^\prime}
      |\vec{p}^{\;\prime}|^2]\frac{1}{|\vec{p}^{\;\prime}|
      \mathcal{N}} \nonumber \\ \zeta_T(Q^2) &= \frac{a_k[(\vec{k}
        \cdot \vec{p}^{\;\prime})^2 - |\vec{k}|^2
        |\vec{p}^{\;\prime}|^2
    ]}{\mathcal{N}|\vec{p}^{\;\prime}||\vec{k} \times
      \vec{p}^{\;\prime}|} \, .
\end{align}     
The above expression can further be simplified as: 
\begin{align}\label{eq:trans_pol_2}
  \mathcal{N} \zeta_L(Q^2) & = \frac{M}{E_{p^\prime}} \big[ a_k E_\nu
    \cos \phi + a_{p^\prime} |\vec{p}^{\;\prime}| \big] \nonumber
  \\ \mathcal{N} \zeta_T(Q^2) & = - a_k \; E_\nu \sin \phi ,
\end{align}
where we have used that $E_\nu = |\vec k|$; $\phi$ is the scattering
angle of the final nucleon with respect to the incident (anti)neutrino
in the LAB frame (see Fig. \ref{fig:pol}).

Explicit expressions for $\zeta_{L,T}$, along with the coefficients
$a_{k, p^\prime}$ and $\mathcal{N}$, are given in
Appendix~\ref{app:pol} in terms of the $\epsilon^{fX}_{ij}$-dependent
FF of Eq. (\ref{eq:NCFFNSI}). Additionally, the ratio of transverse to
longitudinal polarization is obtained and is given in
Eq.~(\ref{eq:ratio}). Such a ratio might be experimentally convenient
because it entails the cancellation of some systematic uncertainties.

\section{Results}
\label{sec:results}

\begin{figure*}
    \centering
    \includegraphics[scale=0.70]{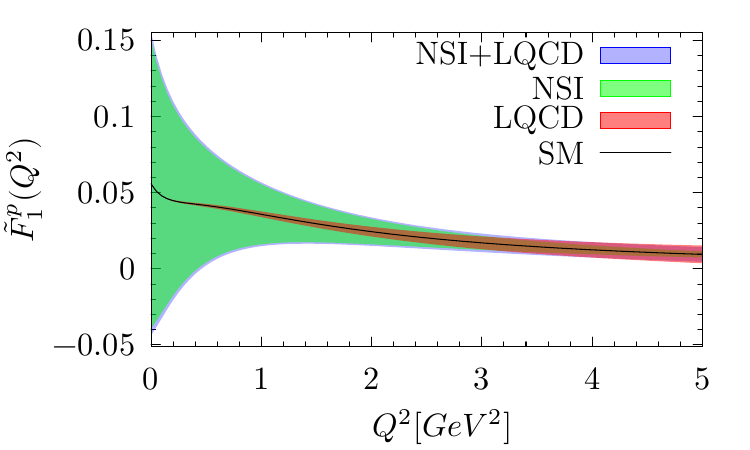}
    \includegraphics[scale=0.70]{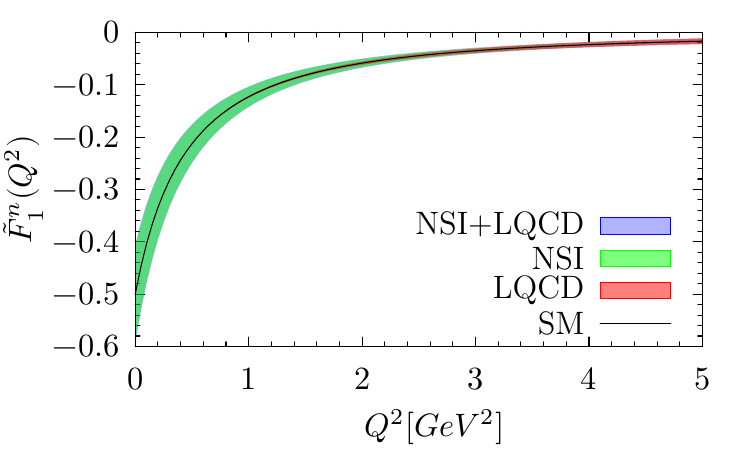}
    \includegraphics[scale=0.70]{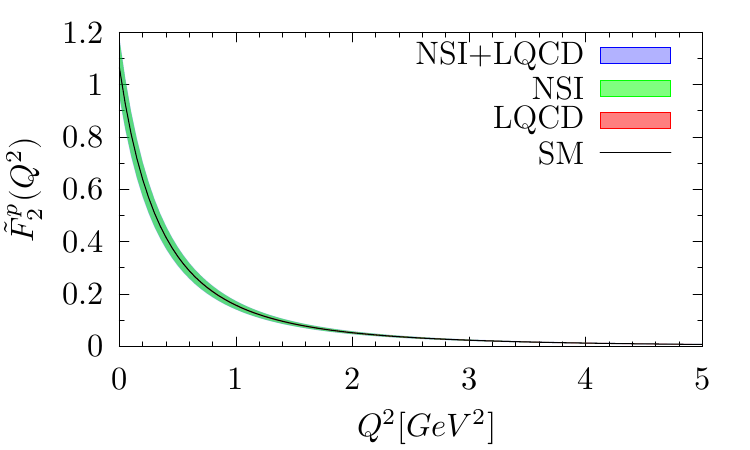}
    \includegraphics[scale=0.70]{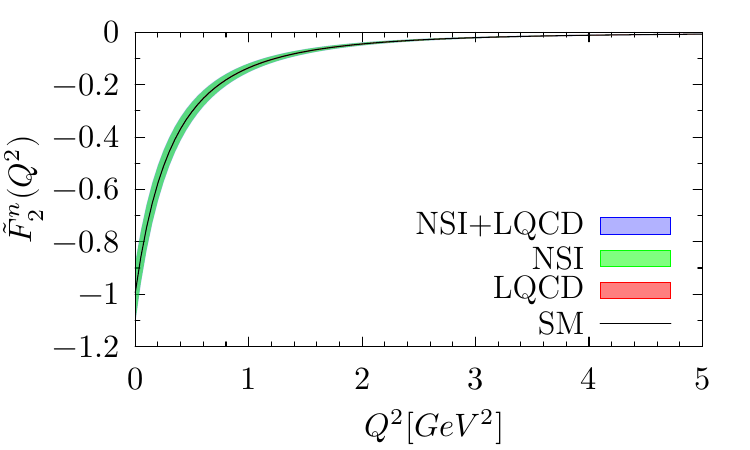}
    \includegraphics[scale=0.70]{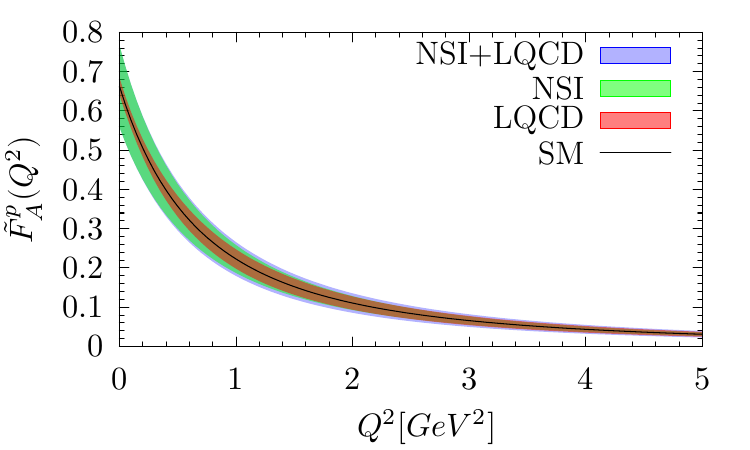}
    \includegraphics[scale=0.70]{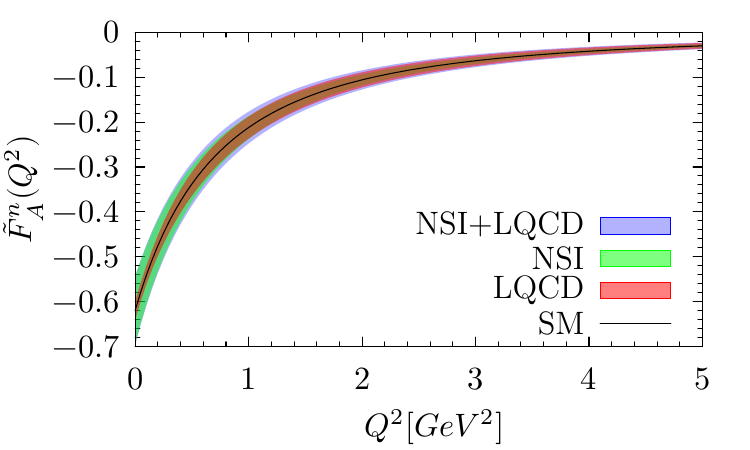}
    \caption{The vector and axial NC form factors defined in
      Eq.~(\ref{eq:NCFFNSI}) are plotted as a function of $Q^2$. The
      black solid lines, representing the SM prediction, are
      accompanied by 90\% CL uncertainty bands. The red innermost ones
      correspond to the errors in the LQCD determination of the axial
      and strange FF by the ETM Collaboration. The green bands show
      the deviations by NSI within the 90\% CL intervals in
      Table~\ref{tab:nsi_para}. The outer blue bands arise from the
      combination of both LQCD and NSI uncertainties. }
    \label{fig:NCformfac}
\end{figure*}

The NC vector and axial FF, $\tilde{F}_{1,2,A}^{p,n}$ introduced in
Eq.~(\ref{eq:NCFFNSI}) are shown in
Fig.~\ref{fig:NCformfac}. Uncertainties in the LQCD input for the
strange vector and flavor-decomposed axial FF, based on the ETM
determination of Refs. \cite{Alexandrou:2019olr,Alexandrou:2021wzv},
together with possible deviations due to NSI, have been displayed in
the form of 90 \% CL bands. It is apparent that the combined
uncertainty is dominated by NSI, except at rather high $Q^2$. In the
vector FF, the small LQCD uncertainty arises solely from
strangeness. The contribution of the latter to $\tilde{F}_{A}^{p,n}$
is also very small, so that the LQCD band width is almost entirely
driven by the uncertainty in the isovector axial FF. Because of the
accidental cancellation among terms in the $(1/2 - 2 s_w^2)$ factor,
$\tilde{F}_{1}^{p}$ is small in the SM and the possible NSI
contribution is relatively large. We should nevertheless warn the
reader about the different abscissa scales in
Fig. \ref{fig:NCformfac}: the NSI band widths are of the same order of
magnitude in all the plots.

\begin{figure*}
    \centering
    \includegraphics[scale=0.7]{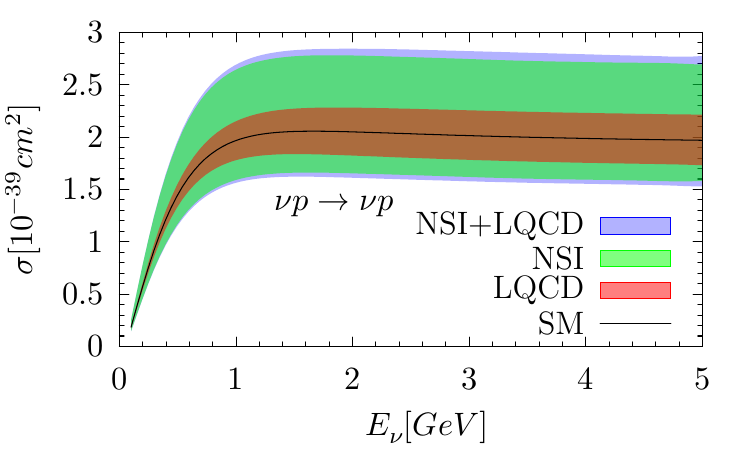}
    \includegraphics[scale=0.7]{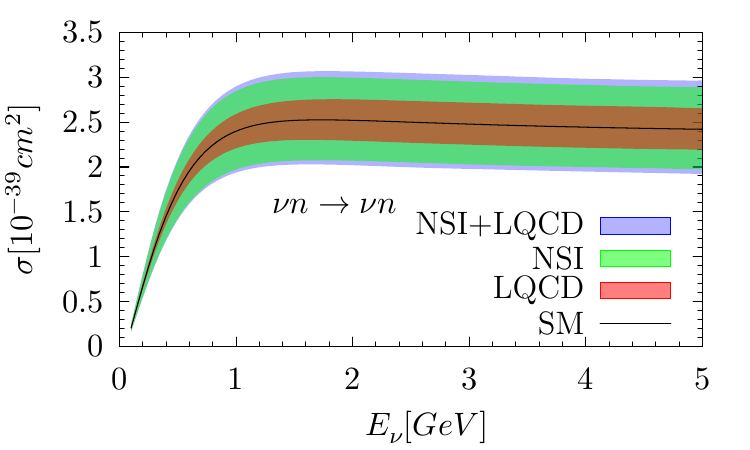}
    \includegraphics[scale=0.7]{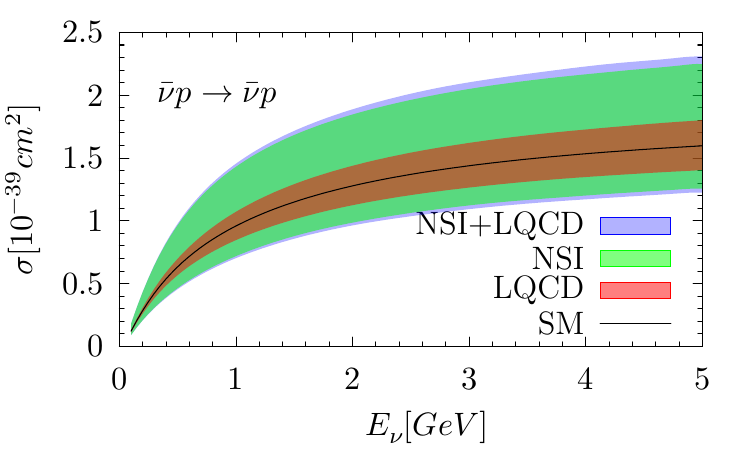}
    \includegraphics[scale=0.7]{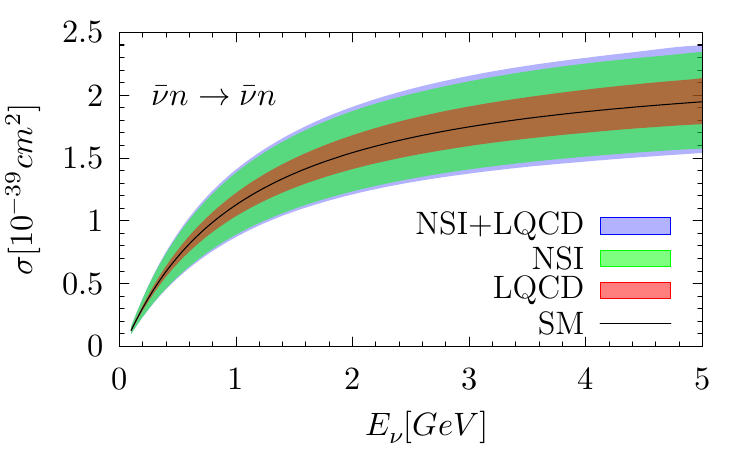}
    
    \caption{Total cross section for NC (anti)neutrino elastic
      scattering as a function of $E_\nu$. The black solid line
      corresponds to the SM. The 90\% CL uncertainty bands from LQCD,
      NSI, and the combination of both, are shown in red, green and
      blue, respectively.  }
       \label{fig:xsec}
\end{figure*}

\begin{figure*}
    \centering
    \includegraphics[scale=0.7]{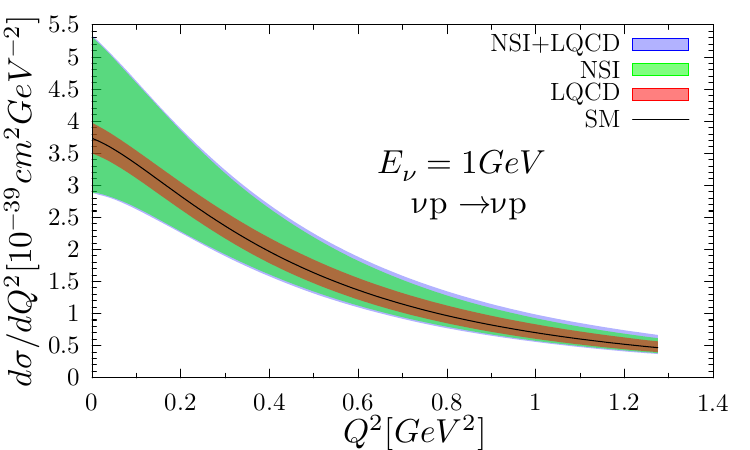}
    \includegraphics[scale=0.7]{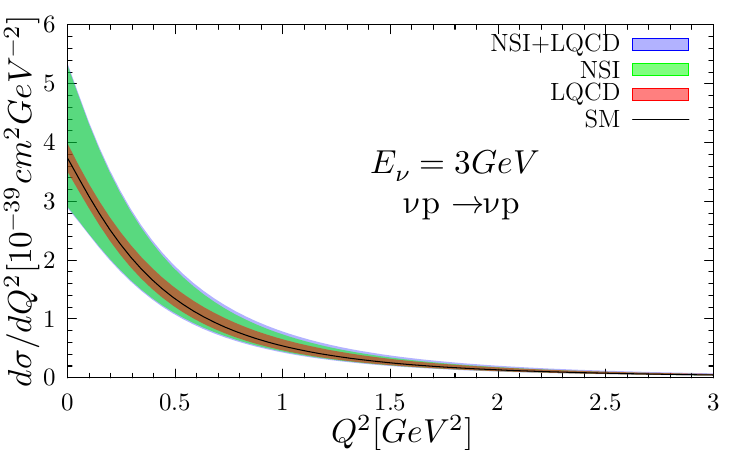}
    \includegraphics[scale=0.7]{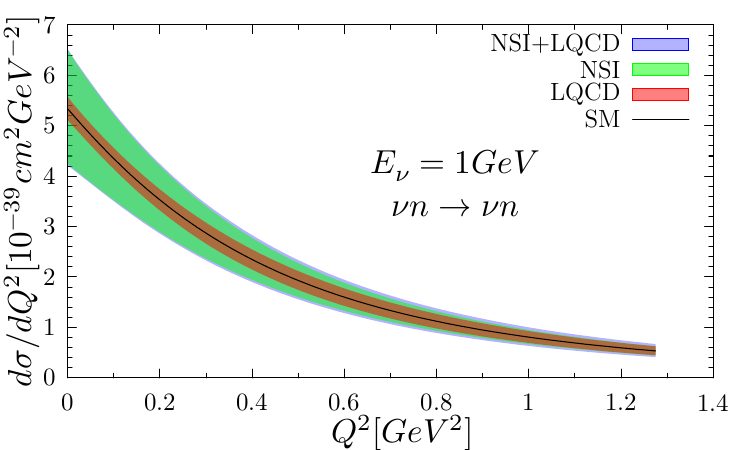}
    \includegraphics[scale=0.7]{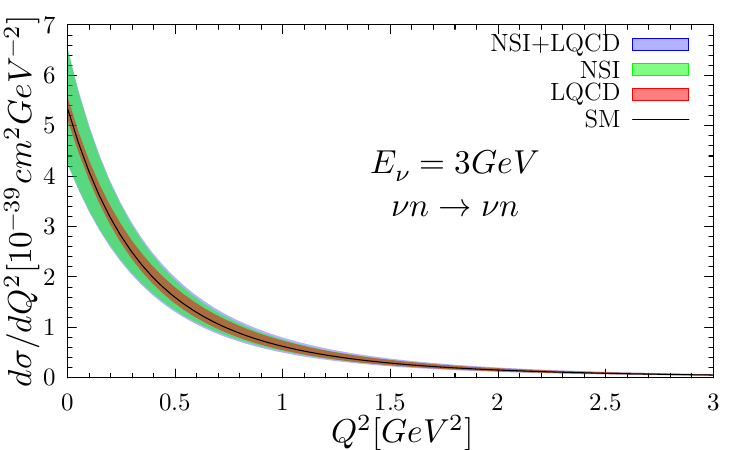}
    \includegraphics[scale=0.7]{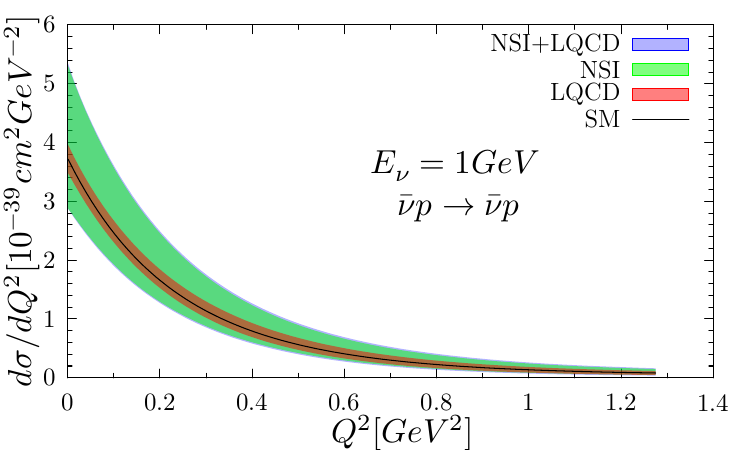}
    \includegraphics[scale=0.7]{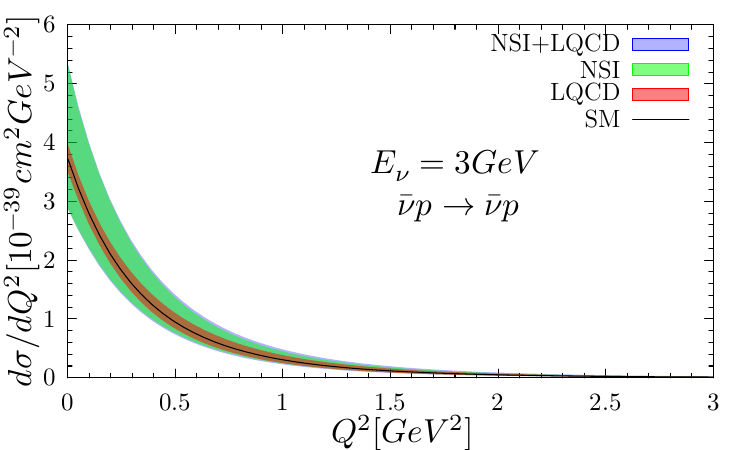}
    \includegraphics[scale=0.7]{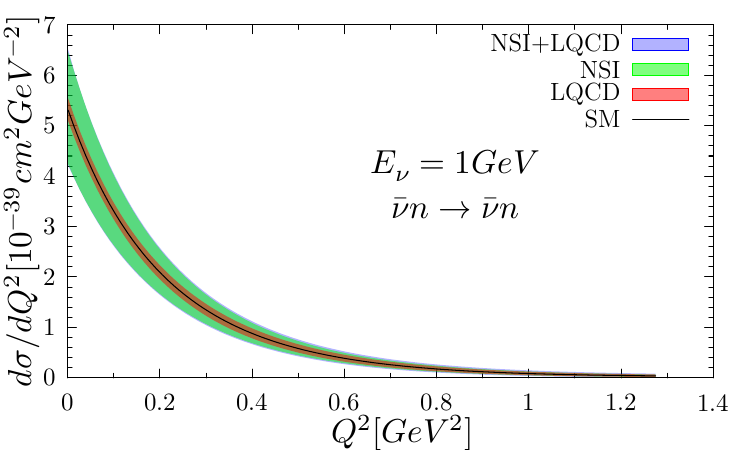}
    \includegraphics[scale=0.7]{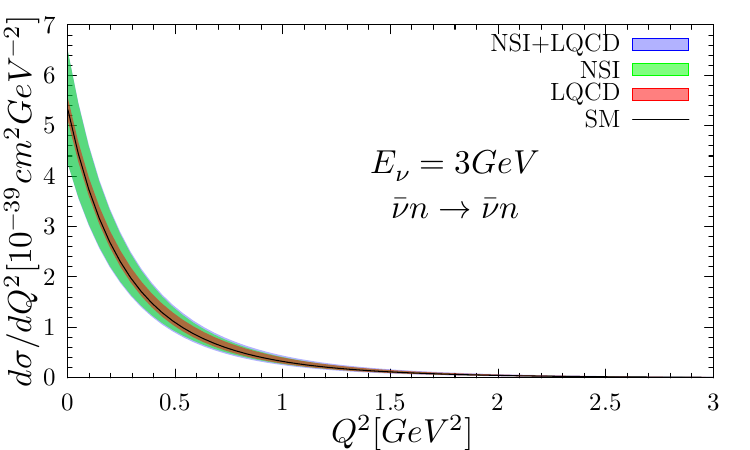}
    \caption{Differential cross section with respect to $Q^2$ for NCE
      channels at $E_\nu=1$ GeV (top) and $3$ GeV (bottom). Lines and
      bands have the same meaning as in Fig.~\ref{fig:xsec}.}
    \label{fig:diffres_xsec_06}
\end{figure*}

Next, we compute the total (integrated) cross-sections for the
processes given in Eq.~(\ref{eq:reac}) to investigate the effects of
NSI. The results are shown in Fig.~\ref{fig:xsec}. The black solid
lines are obtained in the SM, including both vector and axial strange
FF, while the red, green and blue-shaded regions give the possible
deviation due to LQCD, NSI and LQCD plus NSI, respectively. The cross
sections are displayed for incident muon neutrinos and antineutrinos,
as well as for protons and neutron targets. The latter would only be
experimentally accessible with deuteron targets. In this case, the
presence of an additional nucleon would introduce moderate but QCD
dependent corrections~\cite{Singh:1986xh,Shen:2012xz}.

The 90\% bands due to NSI are broader than those from LQCD for the
entire kinematic region. In fact, the blue and green bands largely
overlap. One notices that, given the current bounds, NSI might
ca{\color{red}u}se a significant deviation with respect to the SM
cross section.  For example, at a (anti)neutrino energy of 5 GeV, the
maximal variation for the neutron target is around 20\%, while for the
proton target, it goes as high as 35$\%$ when compared with the SM
predictions. The similar study performed earlier~\cite
{Papoulias:2016edm} was, to the best of our knowledge, the first to
investigate the impact of FSI in the NCE $\overset{(-)}{\nu}$-nucleon
scattering cross section. However, as mentioned after
Eq.~\ref{eq:NCFFNSI}, an ad-hoc $Q^2$ FF was assumed and the isoscalar
FF, $F^{iso}_A$, was also not considered.

Further insight can be gained from differential cross sections. In
Fig.~\ref{fig:diffres_xsec_06}, we plot the $Q^2$ dependence of
$d\sigma/dQ^2$ at two representative (anti)neutrino energies, $E_\nu
=1$ and $3$ GeV. NSI can considerably shift the maximum value of
$d\sigma/dQ^2$, although measurements at low $Q^2$, where such a
maximum is reached, entail experimental difficulties due to the small
nucleon recoil. This is also the region where deuteron corrections are
more pronounced.  The plots reflect the known fact that
$d\sigma/dQ^2(Q^2=0)$ is $E_\nu$ independent and the same for
neutrinos and antineutrinos. See for instance Eqs.~(12,16,18) of
Ref.~\cite{Alvarez-Ruso:2014bla}, whose generalization to the case of
non-zero NSI using Eq.~(\ref{eq:NCFFNSI}) is straightforward. On the
other hand, the target dependence is apparent. At $Q^2=0$, a departure
from SM of up to $+40(-25)\%$ is found for the proton target due to
NSI, while it is around $20\%$ for the neutron target.

\begin{figure*}
    \centering
    \includegraphics[scale=0.7]{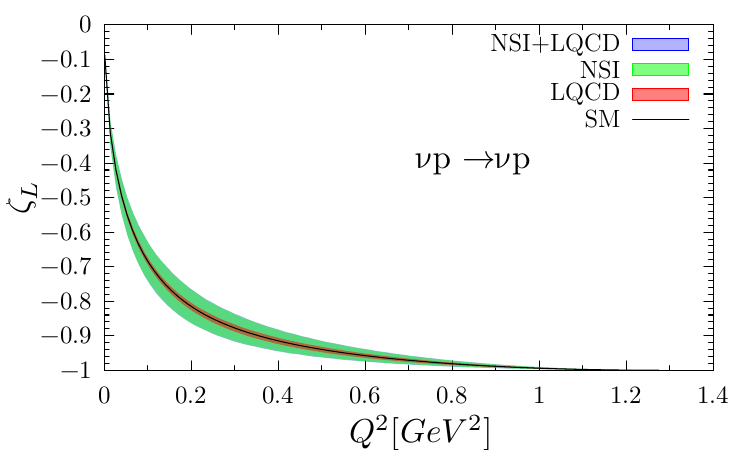}
    \includegraphics[scale=0.7]{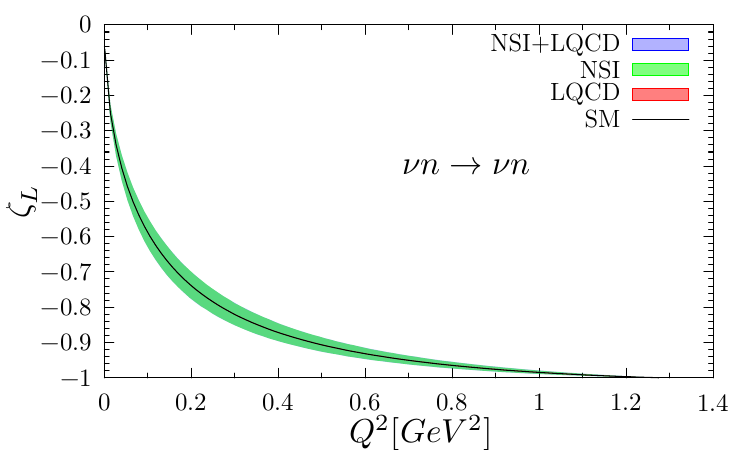}
    \includegraphics[scale=0.7]{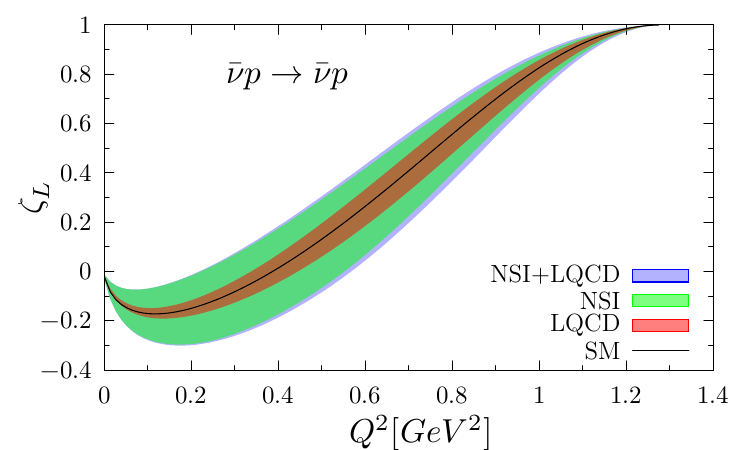}
    \includegraphics[scale=0.7]{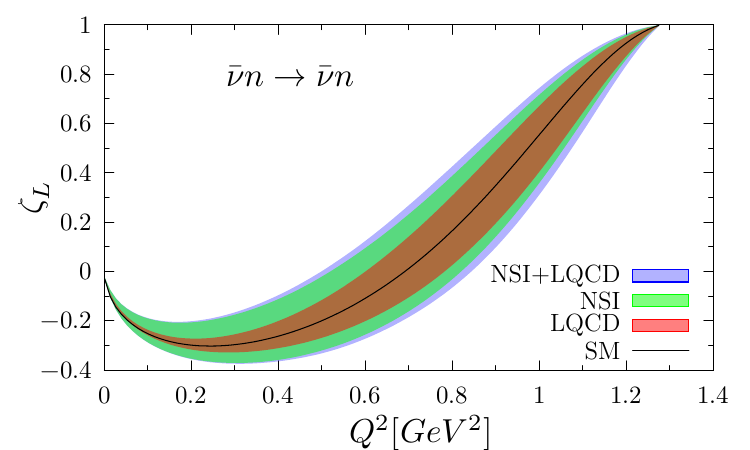}
    \caption{$Q^2$ distribution of the longitudinal polarization of
      the outgoing nucleon, $\zeta_{L}$ for the NCE channels at
      $E_\nu=1$ GeV. Lines and bands have the same meaning as in
      Fig.~\ref{fig:xsec}.}
    \label{fig:diffres_long_06}
\end{figure*}

\begin{figure*}
    \centering
    \includegraphics[scale=0.7]{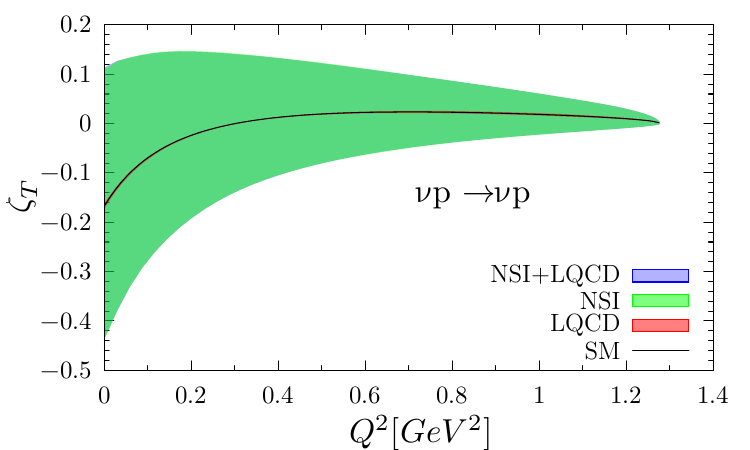}
    \includegraphics[scale=0.7]{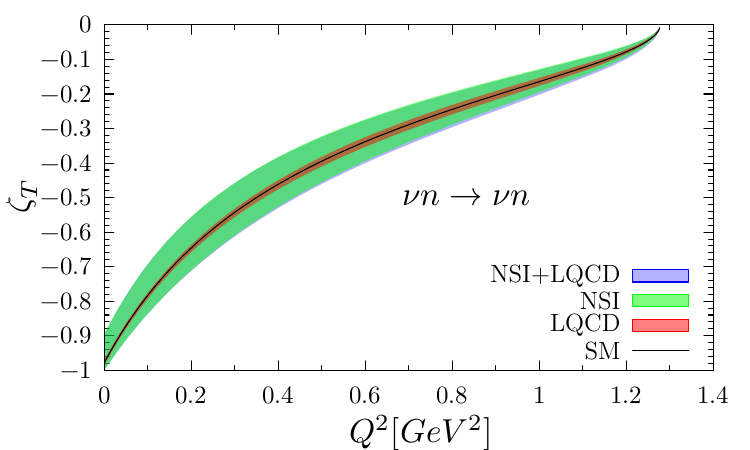}
    \includegraphics[scale=0.7]{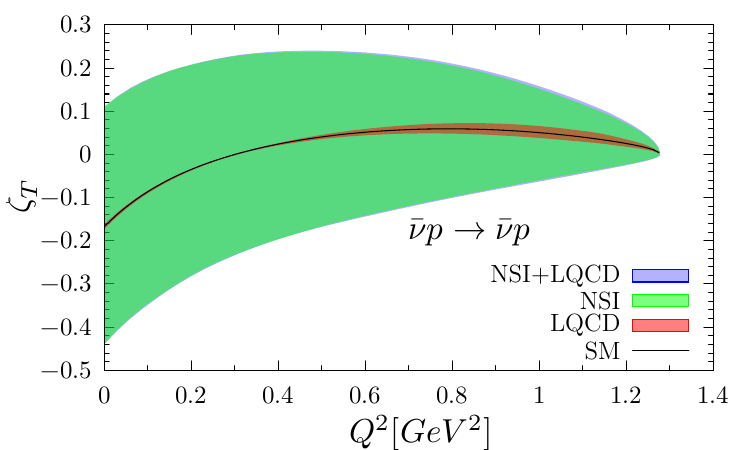}
    \includegraphics[scale=0.7]{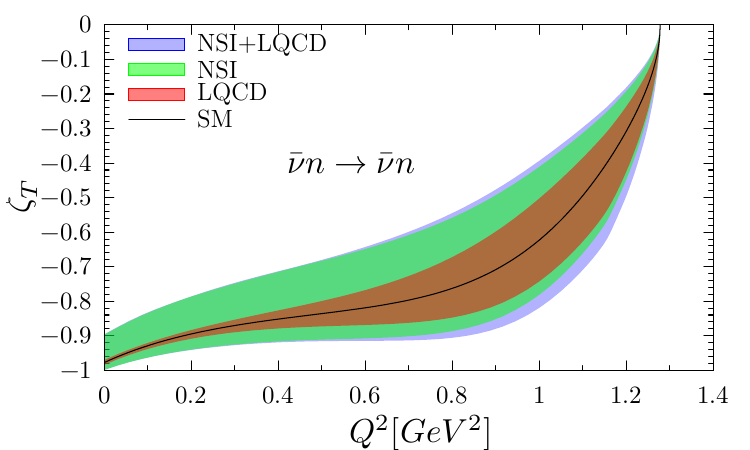}
    \caption{$Q^2$ distribution of the transverse polarization of the
      outgoing nucleon, $\zeta_{T}$ for the NCE channels at $E_\nu=1$
      GeV. Lines and bands have the same meaning as in
      Fig.~\ref{fig:xsec}.}
\label{fig:diffres_long_04}
\end{figure*}

Additional information on the impact of NSI can be obtained from the
polarization of the outgoing nucleon.  We use
Eq.~(\ref{eq:long/transpol}), to obtain the $Q^2$ dependence of
longitudinal ($\zeta_L$) and transverse ($\zeta_T$) polarization. The
explicit expressions for $\zeta_{L,T}$ are given in
Appendix~\ref{app:pol}. The results are shown in
Figs.~\ref{fig:diffres_long_06} and \ref{fig:diffres_long_04} for a
representative (anti)neutrino energy of 1 GeV.

In the case of longitudinal ($\zeta_L$) polarization it is noteworthy
that distinct behaviors are observed for both neutrinos and
antineutrinos. The variation due to LQCD or NSI are nearly negligible
close to both low and high $Q^2$ thresholds, but a significant
variation is evident at intermediate values of $Q^2$, particularly for
antineutrino scattering. The proton polarization in $\bar{\nu}p$
scattering around $Q^2 = 0.2$ GeV displays a relatively large
sensitivity to variations in the NSI couplings compared to the LQCD
uncertainty. At low $Q^2$, the outgoing nucleon's longitudinal
polarization is almost zero and, as $Q^2$ increases, it approaches -1
(1) for the neutrino (antineutrino) channel.\footnote{We note the
different sign of our SM results for the nucleon longitudinal
polarization compared to the ones of Ref.~\cite{Bilenky:2013iua} (for
$\bar{\nu}$ scattering). It is worth stressing that at the largest
$Q^2$, corresponding to backward scattering, the value of $\zeta_L$ is
constrained to be -1 (1) for $\nu (\bar{\nu})$ by the conservation of
the spin projection along the direction of motion.}

On the other hand, in the SM, the transverse component of polarization
of the outgoing proton ($\zeta_T$) is small at $Q^2 \sim 0$ and goes
to zero quite rapidly. Likewise, the LQCD error is almost
negligible. In contrast, we find that NSI bring significant
deviations, specially at low $Q^2$.  Meanwhile, neutrons are found to
be largely polarized in the transverse direction at low $Q^2$, while
this transverse component gradually reaches zero with increasing
$Q^2$. Similar findings were observed previously in
Ref.~\cite{Bilenky:2013iua}. Both the LQCD and NSI bands appear
broader for antineutrinos at middle $Q^2$ ranges.

\section{Summary and Conclusions}
\label{sec:conclusions}

We have studied (anti)neutrino-nucleon NC elastic scattering in
presence of NSI of SM neutrinos with light ($u$, $d$, $s$)
quarks. Assuming a heavy mediator, these interactions take the
well-known form of four-fermion vertices, where the new physics is
concealed in the NSI couplings.  We have therefore derived the NC
matrix elements of the nucleon with NSI, writing them in terms of
vector and axial FF, which encode the strong-interaction physics. We
find that in presence of NSI, these matrix elements depend not only on
the axial isovector FF but also on the isoscalar one, absent in the SM
weak interactions. For this reason, for our numerical predictions we
rely on the flavor decomposition of the axial FF published by the ETM
LQCD collaboration~\cite{Alexandrou:2021wzv}. The availability of
alternative determinations of flavor-diagonal axial form factors would
be important for a better assessment of QCD uncertainties. Under the
assumption that there is no significant new physics in
electron-nucleon scattering, vector FF can be safely obtained from
electron scattering (corrected by radiative corrections). We adopt
this strategy for the vector isovector and isoscalar FF but rely on
the ETM determination~\cite{Alexandrou:2019olr} of the strange
electromagnetic FF.

In our evaluation of scattering observables, only the muon flavor,
with flavor-diagonal, $T$-conserving interactions have been
considered. In this scenario there are four non-zero real couplings
$\epsilon^{uV,dV}_{\mu\mu}$ and $\epsilon^{uA,dA}_{\mu\mu}$, for which
confidence intervals had been obtained from oscillation and
deep-inelastic scattering measurements~\cite{Escrihuela:2011cf}. For
such ranges, we find a significant presence of NSI in differential and
integrated NC (anti)neutrino-nucleon cross sections, particularly for
a proton target and at low $Q^2$. For this choice of target and
kinematics, NSI have a large impact on the polarization of the
outgoing nucleons, especially in the transverse direction.  We have
estimated the uncertainties in the current determination of
(flavor-decomposed) axial and strange vector form factors, finding
that they leave ample room to study NSI.

Neutrino scattering experiments on hydrogen and deuterium are
notoriously challenging. So far, modern neutrino experiments have
chosen heavy targets to maximize statistics. Nevertheless,
(anti)neutrino-nucleon interactions are not only a key ingredient and
a source of uncertainties in theoretical models of
(anti)neutrino-nucleus cross sections but are interesting by
themselves~\cite{Alvarez-Ruso:2022ctb}. The present study adds to this
statement: it shows that, given the current and expected success of
LQCD in obtaining nucleon FF, modern measurements of NC
(anti)neutrino-nucleon elastic scattering can be instrumental in
discovering or constraining non-standard interactions.

\section*{Acknowledgments}
One of the authors, Ilma, greatly acknowledges support from the
University Grants Commission (UGC), India, for providing the Junior
Research Fellowship F.no.~16-9 (June 2019)/2019 (NET/CSIR). This work
has been also partially supported by the Spanish Ministry of Science
and Innovation under Grants No. PID2020-112777GB-I00 and
PID2023-147458NB-C21, funded by MICIU/AEI/10.13039/501100011033, by
the EU STRONG-2020 project under the program H2020-INFRAIA-2018-1,
grant agreement no. 82409, by Generalitat Valenciana grant
CIPROM/2023/59, and by the “Planes Complementarios de I+D+i” program
(grant ASFAE/2022/022) by MICIU with funding from the European Union
NextGenerationEU and Generalitat Valenciana. IRS also acknowledges
support from the Junta de Andalucia (Grant No. FQM-225).

\appendix
\section*{Appendices}
\addcontentsline{toc}{section}{Appendices}
\renewcommand{\thesubsection}{\Alph{subsection}} 
\setcounter{equation}{0}
\renewcommand{\theequation}{A\arabic{equation}}
 
\subsection{Nucleon electromagnetic form factors}
\label{app:ff}

The electromagnetic FF $F^{p,n}_{1,2}(Q^2)$ appearing in
Eq.~(\ref{eq:NCFFNSI}) are generally expressed in terms of Sach's
electric $G_E^{p,n}(Q^2)$ and magnetic $G_M^{p,n}(Q^2)$ FF as:
\begin{align}\label{eq:gmge_ff}
        F_1^N(Q^2) &=\frac{G_E^N(Q^2) +\tau G_M^N(Q^2)}{1+\tau}
        \nonumber \\ F_2^N(Q^2) &=\frac{G_M^N(Q^2)-
          G_E^N(Q^2)}{1+\tau} \quad ; \; N=n,p \; ,
    \end{align}
with $\tau=Q^2/4M^2$. For the latter we rely on the parametrization
by Galster et al.~\cite{Galster:1971kv}:
\begin{align}
     G_E^p &= \left(\frac{1}{1+Q^2/M_V^2}\right)^2 \nonumber \\
        &=\frac{G_M^p}{\mu_p} \; 
         =\frac{G_M^n}{\mu_n} \nonumber \\
        &= -(1+\lambda_n\tau)\frac{G_E^n}{\mu_n\tau}  
\end{align}
where $M_V=0.843$ GeV, $\mu_p=2.792847$, $\mu_n=-1.913043$, and
$\lambda_n=5.6$.

\subsection{Final nucleon polarization in terms of form factors}\label{app:pol}

Besides the incident energy, $E_\nu$, there is only one independent
kinematic variable. We choose $Q^2$ as such a variable but, for the
sake of compactness, in the following expressions we use dimensionless
variables
\begin{align}
    \tau &\equiv \frac{Q^2}{4M^2} \nonumber \\ y &\equiv \frac{p \cdot
      q}{ p \cdot k} = \frac{Q^2}{2 M E_\nu} \,.
\end{align}

The contraction of the leptonic and hadronic tensors, $\mathcal{N}$ of
Eq.~(\ref{eq:N}), is then expressed as:
\begin{align}
    \mathcal{N} &= 2(1-y)\left( \F_A^2 + \frac{\tau
      \G_M^2+\G_E^2}{1+\tau} \right)\nonumber \\ &+ y^2 (\G_M \mp
    \F_A)^2 + \frac{M y}{E_\nu}\left( \F_A^2 - \frac{\tau
      \G_M^2+\G_E^2}{1+\tau} \right) \nonumber \\ &\pm 4y\G_M \F_A \,.
\end{align} 
where (lower )upper sign is for (anti)neutrinos. Equation
~(\ref{eq:gmge_ff}) has been used to revert $\F_{1,2}$ in terms of
$\G_{E,M}$.

The coefficients $a_{k , p^\prime}$ of Eq~(\ref{eq:chi2}) in terms of
these variables are written as:
\begin{align}
    a_k &= \frac{\G_E \, y}{M \tau} \bigg[\pm \G_M y + \F_A(2-y)\bigg]
    \\ a_{p^\prime} &= \frac{1}{2 M \tau (\tau +1)} \bigg[\F_A \G_E
      (y-2) (2 \tau+y)- \F_A \G_M \nonumber \\ &\times (2 \tau+1)
      (y^2+2 \tau ((y-2)y+2)) \nonumber \\ & \mp \bigg( \G_E \G_M y (2
      \tau+y) - (2 \tau+1) (y-2) y \nonumber \\ & \times(\F_A^2
      (\tau+1) +\G_M^2 \tau) \bigg) \bigg] \,.
\end{align}

Substituting them in Eq. (\ref{eq:trans_pol_2}) one obtains for the
polarization vector components along the two orthogonal directions
i.e. transverse ($\zeta_T$) and longitudinal ($\zeta_L$):
\begin{align}
    \mathcal{N} \zeta_T =& - 2 \sin{\phi}[\pm y \G_M +
      (2-y)\F_A]\G_E \label{eq:trans_pol}\\ \mathcal{N} \zeta_L =&
    -\frac{q_0}{\left|\Vec{q} \right|}[\pm y \G_M +(2-y)\F_A]
    \nonumber \\ & \left[(2-y)\G_M\pm y\left( \frac{1+\tau}{\tau}
      \right) \F_A\right] \label{eq:long_pol} \,.
\end{align}
The sinus of the outgoing-nucleon's scattering angle in LAB, with
respect to the incoming (anti)neutrino direction can be expressed as
\begin{align}
\label{eq:sinphi}
\sin{\phi} = \sqrt{ \frac{ 4 \tau ( 1-y ) - y^2}{ 4 \tau ( 1+ \tau) }
} \,.
\end{align}
Finally, the ratio of transverse to longitudinal polarization is given
by
\begin{align}\label{eq:ratio}
\frac{\zeta_T}{\zeta_L} &= \frac{ |\vec{q}| \sin{\phi} }{M} \frac{
  \G_E}{\left[\tau (2-y)\G_M\pm y\left( 1+\tau \right) \F_A\right]}
\,.
\end{align}

\bibliography{biblio}

\begin{thebibliography}{72}%
\makeatletter
\providecommand \@ifxundefined [1]{%
 \@ifx{#1\undefined}
}%
\providecommand \@ifnum [1]{%
 \ifnum #1\expandafter \@firstoftwo
 \else \expandafter \@secondoftwo
 \fi
}%
\providecommand \@ifx [1]{%
 \ifx #1\expandafter \@firstoftwo
 \else \expandafter \@secondoftwo
 \fi
}%
\providecommand \natexlab [1]{#1}%
\providecommand \enquote  [1]{``#1''}%
\providecommand \bibnamefont  [1]{#1}%
\providecommand \bibfnamefont [1]{#1}%
\providecommand \citenamefont [1]{#1}%
\providecommand \href@noop [0]{\@secondoftwo}%
\providecommand \href [0]{\begingroup \@sanitize@url \@href}%
\providecommand \@href[1]{\@@startlink{#1}\@@href}%
\providecommand \@@href[1]{\endgroup#1\@@endlink}%
\providecommand \@sanitize@url [0]{\catcode `\\12\catcode `\$12\catcode
  `\&12\catcode `\#12\catcode `\^12\catcode `\_12\catcode `\%12\relax}%
\providecommand \@@startlink[1]{}%
\providecommand \@@endlink[0]{}%
\providecommand \url  [0]{\begingroup\@sanitize@url \@url }%
\providecommand \@url [1]{\endgroup\@href {#1}{\urlprefix }}%
\providecommand \urlprefix  [0]{URL }%
\providecommand \Eprint [0]{\href }%
\providecommand \doibase [0]{http://dx.doi.org/}%
\providecommand \selectlanguage [0]{\@gobble}%
\providecommand \bibinfo  [0]{\@secondoftwo}%
\providecommand \bibfield  [0]{\@secondoftwo}%
\providecommand \translation [1]{[#1]}%
\providecommand \BibitemOpen [0]{}%
\providecommand \bibitemStop [0]{}%
\providecommand \bibitemNoStop [0]{.\EOS\space}%
\providecommand \EOS [0]{\spacefactor3000\relax}%
\providecommand \BibitemShut  [1]{\csname bibitem#1\endcsname}%
\let\auto@bib@innerbib\@empty
\bibitem [{\citenamefont {Weinberg}(1967)}]{Weinberg:1967tq}%
  \BibitemOpen
  \bibfield  {author} {\bibinfo {author} {\bibfnamefont {S.}~\bibnamefont
  {Weinberg}},\ }\href {\doibase 10.1103/PhysRevLett.19.1264} {\bibfield
  {journal} {\bibinfo  {journal} {Phys. Rev. Lett.}\ }\textbf {\bibinfo
  {volume} {19}},\ \bibinfo {pages} {1264} (\bibinfo {year}
  {1967})}\BibitemShut {NoStop}%
\bibitem [{\citenamefont {Workman}\ \emph {et~al.}(2022)\citenamefont {Workman}
  \emph {et~al.}}]{Workman:2022ynf}%
  \BibitemOpen
  \bibfield  {author} {\bibinfo {author} {\bibfnamefont {R.~L.}\ \bibnamefont
  {Workman}} \emph {et~al.} (\bibinfo {collaboration} {Particle Data Group}),\
  }\href {\doibase 10.1093/ptep/ptac097} {\bibfield  {journal} {\bibinfo
  {journal} {PTEP}\ }\textbf {\bibinfo {volume} {2022}},\ \bibinfo {pages}
  {083C01} (\bibinfo {year} {2022})}\BibitemShut {NoStop}%
\bibitem [{\citenamefont {Lindner}\ \emph {et~al.}(2017)\citenamefont
  {Lindner}, \citenamefont {Rodejohann},\ and\ \citenamefont
  {Xu}}]{Lindner:2016wff}%
  \BibitemOpen
  \bibfield  {author} {\bibinfo {author} {\bibfnamefont {M.}~\bibnamefont
  {Lindner}}, \bibinfo {author} {\bibfnamefont {W.}~\bibnamefont {Rodejohann}},
  \ and\ \bibinfo {author} {\bibfnamefont {X.-J.}\ \bibnamefont {Xu}},\ }\href
  {\doibase 10.1007/JHEP03(2017)097} {\bibfield  {journal} {\bibinfo  {journal}
  {JHEP}\ }\textbf {\bibinfo {volume} {03}},\ \bibinfo {pages} {097} (\bibinfo
  {year} {2017})},\ \Eprint {http://arxiv.org/abs/1612.04150} {arXiv:1612.04150
  [hep-ph]} \BibitemShut {NoStop}%
\bibitem [{\citenamefont {Bischer}\ and\ \citenamefont
  {Rodejohann}(2019)}]{Bischer:2019ttk}%
  \BibitemOpen
  \bibfield  {author} {\bibinfo {author} {\bibfnamefont {I.}~\bibnamefont
  {Bischer}}\ and\ \bibinfo {author} {\bibfnamefont {W.}~\bibnamefont
  {Rodejohann}},\ }\href {\doibase 10.1016/j.nuclphysb.2019.114746} {\bibfield
  {journal} {\bibinfo  {journal} {Nucl. Phys. B}\ }\textbf {\bibinfo {volume}
  {947}},\ \bibinfo {pages} {114746} (\bibinfo {year} {2019})},\ \Eprint
  {http://arxiv.org/abs/1905.08699} {arXiv:1905.08699 [hep-ph]} \BibitemShut
  {NoStop}%
\bibitem [{\citenamefont {Han}\ \emph {et~al.}(2020)\citenamefont {Han},
  \citenamefont {Liao}, \citenamefont {Liu},\ and\ \citenamefont
  {Marfatia}}]{Han:2020pff}%
  \BibitemOpen
  \bibfield  {author} {\bibinfo {author} {\bibfnamefont {T.}~\bibnamefont
  {Han}}, \bibinfo {author} {\bibfnamefont {J.}~\bibnamefont {Liao}}, \bibinfo
  {author} {\bibfnamefont {H.}~\bibnamefont {Liu}}, \ and\ \bibinfo {author}
  {\bibfnamefont {D.}~\bibnamefont {Marfatia}},\ }\href {\doibase
  10.1007/JHEP07(2020)207} {\bibfield  {journal} {\bibinfo  {journal} {JHEP}\
  }\textbf {\bibinfo {volume} {07}},\ \bibinfo {pages} {207} (\bibinfo {year}
  {2020})},\ \Eprint {http://arxiv.org/abs/2004.13869} {arXiv:2004.13869
  [hep-ph]} \BibitemShut {NoStop}%
\bibitem [{\citenamefont {Du}\ \emph {et~al.}(2022)\citenamefont {Du},
  \citenamefont {Li}, \citenamefont {Tang}, \citenamefont {Vihonen},\ and\
  \citenamefont {Yu}}]{Du:2021rdg}%
  \BibitemOpen
  \bibfield  {author} {\bibinfo {author} {\bibfnamefont {Y.}~\bibnamefont
  {Du}}, \bibinfo {author} {\bibfnamefont {H.-L.}\ \bibnamefont {Li}}, \bibinfo
  {author} {\bibfnamefont {J.}~\bibnamefont {Tang}}, \bibinfo {author}
  {\bibfnamefont {S.}~\bibnamefont {Vihonen}}, \ and\ \bibinfo {author}
  {\bibfnamefont {J.-H.}\ \bibnamefont {Yu}},\ }\href {\doibase
  10.1103/PhysRevD.105.075022} {\bibfield  {journal} {\bibinfo  {journal}
  {Phys. Rev. D}\ }\textbf {\bibinfo {volume} {105}},\ \bibinfo {pages}
  {075022} (\bibinfo {year} {2022})},\ \Eprint
  {http://arxiv.org/abs/2106.15800} {arXiv:2106.15800 [hep-ph]} \BibitemShut
  {NoStop}%
\bibitem [{\citenamefont {Wolfenstein}(1978)}]{Wolfenstein:1977ue}%
  \BibitemOpen
  \bibfield  {author} {\bibinfo {author} {\bibfnamefont {L.}~\bibnamefont
  {Wolfenstein}},\ }\href {\doibase 10.1103/PhysRevD.17.2369} {\bibfield
  {journal} {\bibinfo  {journal} {Phys. Rev. D}\ }\textbf {\bibinfo {volume}
  {17}},\ \bibinfo {pages} {2369} (\bibinfo {year} {1978})}\BibitemShut
  {NoStop}%
\bibitem [{\citenamefont {Grossman}(1995)}]{Grossman:1995wx}%
  \BibitemOpen
  \bibfield  {author} {\bibinfo {author} {\bibfnamefont {Y.}~\bibnamefont
  {Grossman}},\ }\href {\doibase 10.1016/0370-2693(95)01069-3} {\bibfield
  {journal} {\bibinfo  {journal} {Phys. Lett. B}\ }\textbf {\bibinfo {volume}
  {359}},\ \bibinfo {pages} {141} (\bibinfo {year} {1995})},\ \Eprint
  {http://arxiv.org/abs/hep-ph/9507344} {arXiv:hep-ph/9507344} \BibitemShut
  {NoStop}%
\bibitem [{\citenamefont {Berezhiani}\ and\ \citenamefont
  {Rossi}(2002)}]{Berezhiani:2001rs}%
  \BibitemOpen
  \bibfield  {author} {\bibinfo {author} {\bibfnamefont {Z.}~\bibnamefont
  {Berezhiani}}\ and\ \bibinfo {author} {\bibfnamefont {A.}~\bibnamefont
  {Rossi}},\ }\href {\doibase 10.1016/S0370-2693(02)01767-7} {\bibfield
  {journal} {\bibinfo  {journal} {Phys. Lett. B}\ }\textbf {\bibinfo {volume}
  {535}},\ \bibinfo {pages} {207} (\bibinfo {year} {2002})},\ \Eprint
  {http://arxiv.org/abs/hep-ph/0111137} {arXiv:hep-ph/0111137} \BibitemShut
  {NoStop}%
\bibitem [{\citenamefont {Davidson}\ \emph {et~al.}(2003)\citenamefont
  {Davidson}, \citenamefont {Pena-Garay}, \citenamefont {Rius},\ and\
  \citenamefont {Santamaria}}]{Davidson:2003ha}%
  \BibitemOpen
  \bibfield  {author} {\bibinfo {author} {\bibfnamefont {S.}~\bibnamefont
  {Davidson}}, \bibinfo {author} {\bibfnamefont {C.}~\bibnamefont
  {Pena-Garay}}, \bibinfo {author} {\bibfnamefont {N.}~\bibnamefont {Rius}}, \
  and\ \bibinfo {author} {\bibfnamefont {A.}~\bibnamefont {Santamaria}},\
  }\href {\doibase 10.1088/1126-6708/2003/03/011} {\bibfield  {journal}
  {\bibinfo  {journal} {JHEP}\ }\textbf {\bibinfo {volume} {03}},\ \bibinfo
  {pages} {011} (\bibinfo {year} {2003})},\ \Eprint
  {http://arxiv.org/abs/hep-ph/0302093} {arXiv:hep-ph/0302093} \BibitemShut
  {NoStop}%
\bibitem [{\citenamefont {Ohlsson}(2013)}]{Ohlsson:2012kf}%
  \BibitemOpen
  \bibfield  {author} {\bibinfo {author} {\bibfnamefont {T.}~\bibnamefont
  {Ohlsson}},\ }\href {\doibase 10.1088/0034-4885/76/4/044201} {\bibfield
  {journal} {\bibinfo  {journal} {Rept. Prog. Phys.}\ }\textbf {\bibinfo
  {volume} {76}},\ \bibinfo {pages} {044201} (\bibinfo {year} {2013})},\
  \Eprint {http://arxiv.org/abs/1209.2710} {arXiv:1209.2710 [hep-ph]}
  \BibitemShut {NoStop}%
\bibitem [{\citenamefont {Farzan}\ and\ \citenamefont
  {Tortola}(2018)}]{Farzan:2017xzy}%
  \BibitemOpen
  \bibfield  {author} {\bibinfo {author} {\bibfnamefont {Y.}~\bibnamefont
  {Farzan}}\ and\ \bibinfo {author} {\bibfnamefont {M.}~\bibnamefont
  {Tortola}},\ }\href {\doibase 10.3389/fphy.2018.00010} {\bibfield  {journal}
  {\bibinfo  {journal} {Front. in Phys.}\ }\textbf {\bibinfo {volume} {6}},\
  \bibinfo {pages} {10} (\bibinfo {year} {2018})},\ \Eprint
  {http://arxiv.org/abs/1710.09360} {arXiv:1710.09360 [hep-ph]} \BibitemShut
  {NoStop}%
\bibitem [{\citenamefont {Masud}\ \emph {et~al.}(2016)\citenamefont {Masud},
  \citenamefont {Chatterjee},\ and\ \citenamefont {Mehta}}]{Masud:2015xva}%
  \BibitemOpen
  \bibfield  {author} {\bibinfo {author} {\bibfnamefont {M.}~\bibnamefont
  {Masud}}, \bibinfo {author} {\bibfnamefont {A.}~\bibnamefont {Chatterjee}}, \
  and\ \bibinfo {author} {\bibfnamefont {P.}~\bibnamefont {Mehta}},\ }\href
  {\doibase 10.1088/0954-3899/43/9/095005/meta} {\bibfield  {journal} {\bibinfo
   {journal} {J. Phys. G}\ }\textbf {\bibinfo {volume} {43}},\ \bibinfo {pages}
  {095005} (\bibinfo {year} {2016})},\ \Eprint
  {http://arxiv.org/abs/1510.08261} {arXiv:1510.08261 [hep-ph]} \BibitemShut
  {NoStop}%
\bibitem [{\citenamefont {de~Gouv\^ea}\ and\ \citenamefont
  {Kelly}(2016)}]{deGouvea:2015ndi}%
  \BibitemOpen
  \bibfield  {author} {\bibinfo {author} {\bibfnamefont {A.}~\bibnamefont
  {de~Gouv\^ea}}\ and\ \bibinfo {author} {\bibfnamefont {K.~J.}\ \bibnamefont
  {Kelly}},\ }\href {\doibase 10.1016/j.nuclphysb.2016.03.013} {\bibfield
  {journal} {\bibinfo  {journal} {Nucl. Phys. B}\ }\textbf {\bibinfo {volume}
  {908}},\ \bibinfo {pages} {318} (\bibinfo {year} {2016})},\ \Eprint
  {http://arxiv.org/abs/1511.05562} {arXiv:1511.05562 [hep-ph]} \BibitemShut
  {NoStop}%
\bibitem [{\citenamefont {Forero}\ and\ \citenamefont
  {Huber}(2016)}]{Forero:2016cmb}%
  \BibitemOpen
  \bibfield  {author} {\bibinfo {author} {\bibfnamefont {D.~V.}\ \bibnamefont
  {Forero}}\ and\ \bibinfo {author} {\bibfnamefont {P.}~\bibnamefont {Huber}},\
  }\href {\doibase 10.1103/PhysRevLett.117.031801} {\bibfield  {journal}
  {\bibinfo  {journal} {Phys. Rev. Lett.}\ }\textbf {\bibinfo {volume} {117}},\
  \bibinfo {pages} {031801} (\bibinfo {year} {2016})},\ \Eprint
  {http://arxiv.org/abs/1601.03736} {arXiv:1601.03736 [hep-ph]} \BibitemShut
  {NoStop}%
\bibitem [{\citenamefont {Esteban}\ \emph {et~al.}(2019)\citenamefont
  {Esteban}, \citenamefont {Gonzalez-Garcia},\ and\ \citenamefont
  {Maltoni}}]{Esteban:2019lfo}%
  \BibitemOpen
  \bibfield  {author} {\bibinfo {author} {\bibfnamefont {I.}~\bibnamefont
  {Esteban}}, \bibinfo {author} {\bibfnamefont {M.~C.}\ \bibnamefont
  {Gonzalez-Garcia}}, \ and\ \bibinfo {author} {\bibfnamefont {M.}~\bibnamefont
  {Maltoni}},\ }\href {\doibase 10.1007/JHEP06(2019)055} {\bibfield  {journal}
  {\bibinfo  {journal} {JHEP}\ }\textbf {\bibinfo {volume} {06}},\ \bibinfo
  {pages} {055} (\bibinfo {year} {2019})},\ \Eprint
  {http://arxiv.org/abs/1905.05203} {arXiv:1905.05203 [hep-ph]} \BibitemShut
  {NoStop}%
\bibitem [{\citenamefont {Capozzi}\ \emph {et~al.}(2020)\citenamefont
  {Capozzi}, \citenamefont {Chatterjee},\ and\ \citenamefont
  {Palazzo}}]{Capozzi:2019iqn}%
  \BibitemOpen
  \bibfield  {author} {\bibinfo {author} {\bibfnamefont {F.}~\bibnamefont
  {Capozzi}}, \bibinfo {author} {\bibfnamefont {S.~S.}\ \bibnamefont
  {Chatterjee}}, \ and\ \bibinfo {author} {\bibfnamefont {A.}~\bibnamefont
  {Palazzo}},\ }\href {\doibase 10.1103/PhysRevLett.124.111801} {\bibfield
  {journal} {\bibinfo  {journal} {Phys. Rev. Lett.}\ }\textbf {\bibinfo
  {volume} {124}},\ \bibinfo {pages} {111801} (\bibinfo {year} {2020})},\
  \Eprint {http://arxiv.org/abs/1908.06992} {arXiv:1908.06992 [hep-ph]}
  \BibitemShut {NoStop}%
\bibitem [{\citenamefont {Barranco}\ \emph {et~al.}(2008)\citenamefont
  {Barranco}, \citenamefont {Miranda}, \citenamefont {Moura},\ and\
  \citenamefont {Valle}}]{Barranco:2007ej}%
  \BibitemOpen
  \bibfield  {author} {\bibinfo {author} {\bibfnamefont {J.}~\bibnamefont
  {Barranco}}, \bibinfo {author} {\bibfnamefont {O.~G.}\ \bibnamefont
  {Miranda}}, \bibinfo {author} {\bibfnamefont {C.~A.}\ \bibnamefont {Moura}},
  \ and\ \bibinfo {author} {\bibfnamefont {J.~W.~F.}\ \bibnamefont {Valle}},\
  }\href {\doibase 10.1103/PhysRevD.77.093014} {\bibfield  {journal} {\bibinfo
  {journal} {Phys. Rev. D}\ }\textbf {\bibinfo {volume} {77}},\ \bibinfo
  {pages} {093014} (\bibinfo {year} {2008})},\ \Eprint
  {http://arxiv.org/abs/0711.0698} {arXiv:0711.0698 [hep-ph]} \BibitemShut
  {NoStop}%
\bibitem [{\citenamefont {Forero}\ and\ \citenamefont
  {Guzzo}(2011)}]{Forero:2011zz}%
  \BibitemOpen
  \bibfield  {author} {\bibinfo {author} {\bibfnamefont {D.~V.}\ \bibnamefont
  {Forero}}\ and\ \bibinfo {author} {\bibfnamefont {M.~M.}\ \bibnamefont
  {Guzzo}},\ }\href {\doibase 10.1103/PhysRevD.84.013002} {\bibfield  {journal}
  {\bibinfo  {journal} {Phys. Rev. D}\ }\textbf {\bibinfo {volume} {84}},\
  \bibinfo {pages} {013002} (\bibinfo {year} {2011})}\BibitemShut {NoStop}%
\bibitem [{\citenamefont {Khan}\ \emph {et~al.}(2014)\citenamefont {Khan},
  \citenamefont {McKay},\ and\ \citenamefont {Tahir}}]{Khan:2014zwa}%
  \BibitemOpen
  \bibfield  {author} {\bibinfo {author} {\bibfnamefont {A.~N.}\ \bibnamefont
  {Khan}}, \bibinfo {author} {\bibfnamefont {D.~W.}\ \bibnamefont {McKay}}, \
  and\ \bibinfo {author} {\bibfnamefont {F.}~\bibnamefont {Tahir}},\ }\href
  {\doibase 10.1103/PhysRevD.90.053008} {\bibfield  {journal} {\bibinfo
  {journal} {Phys. Rev. D}\ }\textbf {\bibinfo {volume} {90}},\ \bibinfo
  {pages} {053008} (\bibinfo {year} {2014})},\ \Eprint
  {http://arxiv.org/abs/1407.4263} {arXiv:1407.4263 [hep-ph]} \BibitemShut
  {NoStop}%
\bibitem [{\citenamefont {Escrihuela}\ \emph {et~al.}(2021)\citenamefont
  {Escrihuela}, \citenamefont {Flores}, \citenamefont {Miranda},\ and\
  \citenamefont {Rend\'on}}]{Escrihuela:2021mud}%
  \BibitemOpen
  \bibfield  {author} {\bibinfo {author} {\bibfnamefont {F.~J.}\ \bibnamefont
  {Escrihuela}}, \bibinfo {author} {\bibfnamefont {L.~J.}\ \bibnamefont
  {Flores}}, \bibinfo {author} {\bibfnamefont {O.~G.}\ \bibnamefont {Miranda}},
  \ and\ \bibinfo {author} {\bibfnamefont {J.}~\bibnamefont {Rend\'on}},\
  }\href {\doibase 10.1007/JHEP07(2021)061} {\bibfield  {journal} {\bibinfo
  {journal} {JHEP}\ }\textbf {\bibinfo {volume} {07}},\ \bibinfo {pages} {061}
  (\bibinfo {year} {2021})},\ \Eprint {http://arxiv.org/abs/2105.06484}
  {arXiv:2105.06484 [hep-ph]} \BibitemShut {NoStop}%
\bibitem [{\citenamefont {Escrihuela}\ \emph {et~al.}(2011)\citenamefont
  {Escrihuela}, \citenamefont {Tortola}, \citenamefont {Valle},\ and\
  \citenamefont {Miranda}}]{Escrihuela:2011cf}%
  \BibitemOpen
  \bibfield  {author} {\bibinfo {author} {\bibfnamefont {F.~J.}\ \bibnamefont
  {Escrihuela}}, \bibinfo {author} {\bibfnamefont {M.}~\bibnamefont {Tortola}},
  \bibinfo {author} {\bibfnamefont {J.~W.~F.}\ \bibnamefont {Valle}}, \ and\
  \bibinfo {author} {\bibfnamefont {O.~G.}\ \bibnamefont {Miranda}},\ }\href
  {\doibase 10.1103/PhysRevD.83.093002} {\bibfield  {journal} {\bibinfo
  {journal} {Phys. Rev. D}\ }\textbf {\bibinfo {volume} {83}},\ \bibinfo
  {pages} {093002} (\bibinfo {year} {2011})},\ \Eprint
  {http://arxiv.org/abs/1103.1366} {arXiv:1103.1366 [hep-ph]} \BibitemShut
  {NoStop}%
\bibitem [{\citenamefont {Akimov}\ \emph {et~al.}(2017)\citenamefont {Akimov}
  \emph {et~al.}}]{COHERENT:2017ipa}%
  \BibitemOpen
  \bibfield  {author} {\bibinfo {author} {\bibfnamefont {D.}~\bibnamefont
  {Akimov}} \emph {et~al.} (\bibinfo {collaboration} {COHERENT}),\ }\href
  {\doibase 10.1126/science.aao0990} {\bibfield  {journal} {\bibinfo  {journal}
  {Science}\ }\textbf {\bibinfo {volume} {357}},\ \bibinfo {pages} {1123}
  (\bibinfo {year} {2017})},\ \Eprint {http://arxiv.org/abs/1708.01294}
  {arXiv:1708.01294 [nucl-ex]} \BibitemShut {NoStop}%
\bibitem [{\citenamefont {Coloma}\ \emph {et~al.}(2017)\citenamefont {Coloma},
  \citenamefont {Gonzalez-Garcia}, \citenamefont {Maltoni},\ and\ \citenamefont
  {Schwetz}}]{Coloma:2017ncl}%
  \BibitemOpen
  \bibfield  {author} {\bibinfo {author} {\bibfnamefont {P.}~\bibnamefont
  {Coloma}}, \bibinfo {author} {\bibfnamefont {M.~C.}\ \bibnamefont
  {Gonzalez-Garcia}}, \bibinfo {author} {\bibfnamefont {M.}~\bibnamefont
  {Maltoni}}, \ and\ \bibinfo {author} {\bibfnamefont {T.}~\bibnamefont
  {Schwetz}},\ }\href {\doibase 10.1103/PhysRevD.96.115007} {\bibfield
  {journal} {\bibinfo  {journal} {Phys. Rev. D}\ }\textbf {\bibinfo {volume}
  {96}},\ \bibinfo {pages} {115007} (\bibinfo {year} {2017})},\ \Eprint
  {http://arxiv.org/abs/1708.02899} {arXiv:1708.02899 [hep-ph]} \BibitemShut
  {NoStop}%
\bibitem [{\citenamefont {Liao}\ and\ \citenamefont
  {Marfatia}(2017)}]{Liao:2017uzy}%
  \BibitemOpen
  \bibfield  {author} {\bibinfo {author} {\bibfnamefont {J.}~\bibnamefont
  {Liao}}\ and\ \bibinfo {author} {\bibfnamefont {D.}~\bibnamefont
  {Marfatia}},\ }\href {\doibase 10.1016/j.physletb.2017.10.046} {\bibfield
  {journal} {\bibinfo  {journal} {Phys. Lett. B}\ }\textbf {\bibinfo {volume}
  {775}},\ \bibinfo {pages} {54} (\bibinfo {year} {2017})},\ \Eprint
  {http://arxiv.org/abs/1708.04255} {arXiv:1708.04255 [hep-ph]} \BibitemShut
  {NoStop}%
\bibitem [{\citenamefont {Aristizabal~Sierra}\ \emph
  {et~al.}(2018)\citenamefont {Aristizabal~Sierra}, \citenamefont {De~Romeri},\
  and\ \citenamefont {Rojas}}]{AristizabalSierra:2018eqm}%
  \BibitemOpen
  \bibfield  {author} {\bibinfo {author} {\bibfnamefont {D.}~\bibnamefont
  {Aristizabal~Sierra}}, \bibinfo {author} {\bibfnamefont {V.}~\bibnamefont
  {De~Romeri}}, \ and\ \bibinfo {author} {\bibfnamefont {N.}~\bibnamefont
  {Rojas}},\ }\href {\doibase 10.1103/PhysRevD.98.075018} {\bibfield  {journal}
  {\bibinfo  {journal} {Phys. Rev. D}\ }\textbf {\bibinfo {volume} {98}},\
  \bibinfo {pages} {075018} (\bibinfo {year} {2018})},\ \Eprint
  {http://arxiv.org/abs/1806.07424} {arXiv:1806.07424 [hep-ph]} \BibitemShut
  {NoStop}%
\bibitem [{\citenamefont {Denton}\ and\ \citenamefont
  {Gehrlein}(2021)}]{Denton:2020hop}%
  \BibitemOpen
  \bibfield  {author} {\bibinfo {author} {\bibfnamefont {P.~B.}\ \bibnamefont
  {Denton}}\ and\ \bibinfo {author} {\bibfnamefont {J.}~\bibnamefont
  {Gehrlein}},\ }\href {\doibase 10.1007/JHEP04(2021)266} {\bibfield  {journal}
  {\bibinfo  {journal} {JHEP}\ }\textbf {\bibinfo {volume} {04}},\ \bibinfo
  {pages} {266} (\bibinfo {year} {2021})},\ \Eprint
  {http://arxiv.org/abs/2008.06062} {arXiv:2008.06062 [hep-ph]} \BibitemShut
  {NoStop}%
\bibitem [{\citenamefont {Khan}\ \emph {et~al.}(2021)\citenamefont {Khan},
  \citenamefont {McKay},\ and\ \citenamefont {Rodejohann}}]{Khan:2021wzy}%
  \BibitemOpen
  \bibfield  {author} {\bibinfo {author} {\bibfnamefont {A.~N.}\ \bibnamefont
  {Khan}}, \bibinfo {author} {\bibfnamefont {D.~W.}\ \bibnamefont {McKay}}, \
  and\ \bibinfo {author} {\bibfnamefont {W.}~\bibnamefont {Rodejohann}},\
  }\href {\doibase 10.1103/PhysRevD.104.015019} {\bibfield  {journal} {\bibinfo
   {journal} {Phys. Rev. D}\ }\textbf {\bibinfo {volume} {104}},\ \bibinfo
  {pages} {015019} (\bibinfo {year} {2021})},\ \Eprint
  {http://arxiv.org/abs/2104.00425} {arXiv:2104.00425 [hep-ph]} \BibitemShut
  {NoStop}%
\bibitem [{\citenamefont {Flores}\ \emph {et~al.}(2022)\citenamefont {Flores},
  \citenamefont {Nath},\ and\ \citenamefont {Peinado}}]{Flores:2021kzl}%
  \BibitemOpen
  \bibfield  {author} {\bibinfo {author} {\bibfnamefont {L.~J.}\ \bibnamefont
  {Flores}}, \bibinfo {author} {\bibfnamefont {N.}~\bibnamefont {Nath}}, \ and\
  \bibinfo {author} {\bibfnamefont {E.}~\bibnamefont {Peinado}},\ }\href
  {\doibase 10.1103/PhysRevD.105.055010} {\bibfield  {journal} {\bibinfo
  {journal} {Phys. Rev. D}\ }\textbf {\bibinfo {volume} {105}},\ \bibinfo
  {pages} {055010} (\bibinfo {year} {2022})},\ \Eprint
  {http://arxiv.org/abs/2112.05103} {arXiv:2112.05103 [hep-ph]} \BibitemShut
  {NoStop}%
\bibitem [{\citenamefont {De~Romeri}\ \emph {et~al.}(2023)\citenamefont
  {De~Romeri}, \citenamefont {Miranda}, \citenamefont {Papoulias},
  \citenamefont {Sanchez~Garcia}, \citenamefont {T\'ortola},\ and\
  \citenamefont {Valle}}]{DeRomeri:2022twg}%
  \BibitemOpen
  \bibfield  {author} {\bibinfo {author} {\bibfnamefont {V.}~\bibnamefont
  {De~Romeri}}, \bibinfo {author} {\bibfnamefont {O.~G.}\ \bibnamefont
  {Miranda}}, \bibinfo {author} {\bibfnamefont {D.~K.}\ \bibnamefont
  {Papoulias}}, \bibinfo {author} {\bibfnamefont {G.}~\bibnamefont
  {Sanchez~Garcia}}, \bibinfo {author} {\bibfnamefont {M.}~\bibnamefont
  {T\'ortola}}, \ and\ \bibinfo {author} {\bibfnamefont {J.~W.~F.}\
  \bibnamefont {Valle}},\ }\href {\doibase 10.1007/JHEP04(2023)035} {\bibfield
  {journal} {\bibinfo  {journal} {JHEP}\ }\textbf {\bibinfo {volume} {04}},\
  \bibinfo {pages} {035} (\bibinfo {year} {2023})},\ \Eprint
  {http://arxiv.org/abs/2211.11905} {arXiv:2211.11905 [hep-ph]} \BibitemShut
  {NoStop}%
\bibitem [{\citenamefont {Papoulias}\ and\ \citenamefont
  {Kosmas}(2016)}]{Papoulias:2016edm}%
  \BibitemOpen
  \bibfield  {author} {\bibinfo {author} {\bibfnamefont {D.~K.}\ \bibnamefont
  {Papoulias}}\ and\ \bibinfo {author} {\bibfnamefont {T.~S.}\ \bibnamefont
  {Kosmas}},\ }\href {\doibase 10.1155/2016/1490860} {\bibfield  {journal}
  {\bibinfo  {journal} {Adv. High Energy Phys.}\ }\textbf {\bibinfo {volume}
  {2016}},\ \bibinfo {pages} {1490860} (\bibinfo {year} {2016})},\ \Eprint
  {http://arxiv.org/abs/1611.05069} {arXiv:1611.05069 [hep-ph]} \BibitemShut
  {NoStop}%
\bibitem [{\citenamefont {Borah}\ \emph {et~al.}(2024)\citenamefont {Borah},
  \citenamefont {Betancourt}, \citenamefont {Hill}, \citenamefont {Junk},\ and\
  \citenamefont {Tomalak}}]{Borah:2024hvo}%
  \BibitemOpen
  \bibfield  {author} {\bibinfo {author} {\bibfnamefont {K.}~\bibnamefont
  {Borah}}, \bibinfo {author} {\bibfnamefont {M.}~\bibnamefont {Betancourt}},
  \bibinfo {author} {\bibfnamefont {R.~J.}\ \bibnamefont {Hill}}, \bibinfo
  {author} {\bibfnamefont {T.}~\bibnamefont {Junk}}, \ and\ \bibinfo {author}
  {\bibfnamefont {O.}~\bibnamefont {Tomalak}},\ }\href {\doibase
  10.1103/PhysRevD.110.013004} {\bibfield  {journal} {\bibinfo  {journal}
  {Phys. Rev. D}\ }\textbf {\bibinfo {volume} {110}},\ \bibinfo {pages}
  {013004} (\bibinfo {year} {2024})},\ \Eprint
  {http://arxiv.org/abs/2403.04687} {arXiv:2403.04687 [hep-ph]} \BibitemShut
  {NoStop}%
\bibitem [{\citenamefont {Tomalak}\ \emph {et~al.}(2024)\citenamefont
  {Tomalak}, \citenamefont {Betancourt}, \citenamefont {Borah}, \citenamefont
  {Hill},\ and\ \citenamefont {Junk}}]{Tomalak:2024yvq}%
  \BibitemOpen
  \bibfield  {author} {\bibinfo {author} {\bibfnamefont {O.}~\bibnamefont
  {Tomalak}}, \bibinfo {author} {\bibfnamefont {M.}~\bibnamefont {Betancourt}},
  \bibinfo {author} {\bibfnamefont {K.}~\bibnamefont {Borah}}, \bibinfo
  {author} {\bibfnamefont {R.~J.}\ \bibnamefont {Hill}}, \ and\ \bibinfo
  {author} {\bibfnamefont {T.}~\bibnamefont {Junk}},\ }\href {\doibase
  10.1016/j.physletb.2024.138718} {\bibfield  {journal} {\bibinfo  {journal}
  {Phys. Lett. B}\ }\textbf {\bibinfo {volume} {854}},\ \bibinfo {pages}
  {138718} (\bibinfo {year} {2024})},\ \Eprint
  {http://arxiv.org/abs/2402.14115} {arXiv:2402.14115 [hep-ph]} \BibitemShut
  {NoStop}%
\bibitem [{\citenamefont {Meyer}\ \emph {et~al.}(2022)\citenamefont {Meyer},
  \citenamefont {Walker-Loud},\ and\ \citenamefont
  {Wilkinson}}]{Meyer:2022mix}%
  \BibitemOpen
  \bibfield  {author} {\bibinfo {author} {\bibfnamefont {A.~S.}\ \bibnamefont
  {Meyer}}, \bibinfo {author} {\bibfnamefont {A.}~\bibnamefont {Walker-Loud}},
  \ and\ \bibinfo {author} {\bibfnamefont {C.}~\bibnamefont {Wilkinson}},\
  }\href {\doibase 10.1146/annurev-nucl-010622-120608} {\bibfield  {journal}
  {\bibinfo  {journal} {Ann. Rev. Nucl. Part. Sci.}\ }\textbf {\bibinfo
  {volume} {72}},\ \bibinfo {pages} {205} (\bibinfo {year} {2022})},\ \Eprint
  {http://arxiv.org/abs/2201.01839} {arXiv:2201.01839 [hep-lat]} \BibitemShut
  {NoStop}%
\bibitem [{\citenamefont {Alvarez-Ruso}\ \emph {et~al.}(2022)\citenamefont
  {Alvarez-Ruso} \emph {et~al.}}]{Alvarez-Ruso:2022ctb}%
  \BibitemOpen
  \bibfield  {author} {\bibinfo {author} {\bibfnamefont {L.}~\bibnamefont
  {Alvarez-Ruso}} \emph {et~al.},\ }\href@noop {} {\  (\bibinfo {year}
  {2022})},\ \Eprint {http://arxiv.org/abs/2203.11298} {arXiv:2203.11298
  [hep-ex]} \BibitemShut {NoStop}%
\bibitem [{\citenamefont {Duyang}\ \emph {et~al.}(2024)\citenamefont {Duyang},
  \citenamefont {Guo}, \citenamefont {Mishra},\ and\ \citenamefont
  {Petti}}]{Duyang:2024ucj}%
  \BibitemOpen
  \bibfield  {author} {\bibinfo {author} {\bibfnamefont {H.}~\bibnamefont
  {Duyang}}, \bibinfo {author} {\bibfnamefont {B.}~\bibnamefont {Guo}},
  \bibinfo {author} {\bibfnamefont {S.~R.}\ \bibnamefont {Mishra}}, \ and\
  \bibinfo {author} {\bibfnamefont {R.}~\bibnamefont {Petti}},\ }\href
  {\doibase 10.1140/epjp/s13360-024-05783-y} {\bibfield  {journal} {\bibinfo
  {journal} {Eur. Phys. J. Plus}\ }\textbf {\bibinfo {volume} {139}},\ \bibinfo
  {pages} {1014} (\bibinfo {year} {2024})}\BibitemShut {NoStop}%
\bibitem [{\citenamefont {Cai}\ \emph {et~al.}(2023)\citenamefont {Cai} \emph
  {et~al.}}]{MINERvA:2023avz}%
  \BibitemOpen
  \bibfield  {author} {\bibinfo {author} {\bibfnamefont {T.}~\bibnamefont
  {Cai}} \emph {et~al.} (\bibinfo {collaboration} {MINERvA}),\ }\href {\doibase
  10.1038/s41586-022-05478-3} {\bibfield  {journal} {\bibinfo  {journal}
  {Nature}\ }\textbf {\bibinfo {volume} {614}},\ \bibinfo {pages} {48}
  (\bibinfo {year} {2023})}\BibitemShut {NoStop}%
\bibitem [{\citenamefont {Kopp}\ \emph {et~al.}(2024)\citenamefont {Kopp},
  \citenamefont {Rocco},\ and\ \citenamefont {Tabrizi}}]{Kopp:2024yvh}%
  \BibitemOpen
  \bibfield  {author} {\bibinfo {author} {\bibfnamefont {J.}~\bibnamefont
  {Kopp}}, \bibinfo {author} {\bibfnamefont {N.}~\bibnamefont {Rocco}}, \ and\
  \bibinfo {author} {\bibfnamefont {Z.}~\bibnamefont {Tabrizi}},\ }\href
  {\doibase 10.1007/JHEP08(2024)187} {\bibfield  {journal} {\bibinfo  {journal}
  {JHEP}\ }\textbf {\bibinfo {volume} {08}},\ \bibinfo {pages} {187} (\bibinfo
  {year} {2024})},\ \Eprint {http://arxiv.org/abs/2401.07902} {arXiv:2401.07902
  [hep-ph]} \BibitemShut {NoStop}%
\bibitem [{\citenamefont {Kong}\ \emph {et~al.}(2023)\citenamefont {Kong},
  \citenamefont {Lai}, \citenamefont {Li}, \citenamefont {Yan},\ and\
  \citenamefont {Yang}}]{Kong:2023kkd}%
  \BibitemOpen
  \bibfield  {author} {\bibinfo {author} {\bibfnamefont {Y.-R.}\ \bibnamefont
  {Kong}}, \bibinfo {author} {\bibfnamefont {L.-F.}\ \bibnamefont {Lai}},
  \bibinfo {author} {\bibfnamefont {X.-Q.}\ \bibnamefont {Li}}, \bibinfo
  {author} {\bibfnamefont {X.-S.}\ \bibnamefont {Yan}}, \ and\ \bibinfo
  {author} {\bibfnamefont {Y.-D.}\ \bibnamefont {Yang}},\ }\href@noop {} {\
  (\bibinfo {year} {2023})},\ \Eprint {http://arxiv.org/abs/2307.07239}
  {arXiv:2307.07239 [hep-ph]} \BibitemShut {NoStop}%
\bibitem [{\citenamefont {Hagiwara}\ \emph {et~al.}(2003)\citenamefont
  {Hagiwara}, \citenamefont {Mawatari},\ and\ \citenamefont
  {Yokoya}}]{Hagiwara:2003di}%
  \BibitemOpen
  \bibfield  {author} {\bibinfo {author} {\bibfnamefont {K.}~\bibnamefont
  {Hagiwara}}, \bibinfo {author} {\bibfnamefont {K.}~\bibnamefont {Mawatari}},
  \ and\ \bibinfo {author} {\bibfnamefont {H.}~\bibnamefont {Yokoya}},\ }\href
  {\doibase 10.1016/S0550-3213(03)00575-3} {\bibfield  {journal} {\bibinfo
  {journal} {Nucl. Phys.}\ }\textbf {\bibinfo {volume} {B668}},\ \bibinfo
  {pages} {364} (\bibinfo {year} {2003})},\ \bibinfo {note} {[Erratum: Nucl.
  Phys.B701,405(2004)]},\ \Eprint {http://arxiv.org/abs/hep-ph/0305324}
  {arXiv:hep-ph/0305324 [hep-ph]} \BibitemShut {NoStop}%
\bibitem [{\citenamefont {Graczyk}(2005)}]{Graczyk:2004uy}%
  \BibitemOpen
  \bibfield  {author} {\bibinfo {author} {\bibfnamefont {K.~M.}\ \bibnamefont
  {Graczyk}},\ }\href {\doibase 10.1016/j.nuclphysa.2004.10.029} {\bibfield
  {journal} {\bibinfo  {journal} {Nucl. Phys.}\ }\textbf {\bibinfo {volume}
  {A748}},\ \bibinfo {pages} {313} (\bibinfo {year} {2005})},\ \Eprint
  {http://arxiv.org/abs/hep-ph/0407275} {arXiv:hep-ph/0407275 [hep-ph]}
  \BibitemShut {NoStop}%
\bibitem [{\citenamefont {Kuzmin}\ \emph
  {et~al.}(2005{\natexlab{a}})\citenamefont {Kuzmin}, \citenamefont
  {Lyubushkin},\ and\ \citenamefont {Naumov}}]{Kuzmin:2004yb}%
  \BibitemOpen
  \bibfield  {author} {\bibinfo {author} {\bibfnamefont {K.~S.}\ \bibnamefont
  {Kuzmin}}, \bibinfo {author} {\bibfnamefont {V.~V.}\ \bibnamefont
  {Lyubushkin}}, \ and\ \bibinfo {author} {\bibfnamefont {V.~A.}\ \bibnamefont
  {Naumov}},\ }\bibfield  {booktitle} {\emph {\bibinfo {booktitle}
  {{Proceedings, 3rd International Workshop on Neutrino-nucleus interactions in
  the few GeV region (NUINT 04): Assergi, Italy, March 17-21, 2004}}},\ }\href
  {\doibase 10.1016/j.nuclphysbps.2004.11.221} {\bibfield  {journal} {\bibinfo
  {journal} {Nucl. Phys. Proc. Suppl.}\ }\textbf {\bibinfo {volume} {139}},\
  \bibinfo {pages} {154} (\bibinfo {year} {2005}{\natexlab{a}})},\ \bibinfo
  {note} {[,154(2004)]},\ \Eprint {http://arxiv.org/abs/hep-ph/0408107}
  {arXiv:hep-ph/0408107 [hep-ph]} \BibitemShut {NoStop}%
\bibitem [{\citenamefont {Kuzmin}\ \emph
  {et~al.}(2005{\natexlab{b}})\citenamefont {Kuzmin}, \citenamefont
  {Lyubushkin},\ and\ \citenamefont {Naumov}}]{Kuzmin:2004ya}%
  \BibitemOpen
  \bibfield  {author} {\bibinfo {author} {\bibfnamefont {K.~S.}\ \bibnamefont
  {Kuzmin}}, \bibinfo {author} {\bibfnamefont {V.~V.}\ \bibnamefont
  {Lyubushkin}}, \ and\ \bibinfo {author} {\bibfnamefont {V.~A.}\ \bibnamefont
  {Naumov}},\ }\bibfield  {booktitle} {\emph {\bibinfo {booktitle}
  {{Proceedings, 3rd International Workshop on Neutrino-nucleus interactions in
  the few GeV region (NUINT 04): Assergi, Italy, March 17-21, 2004}}},\ }\href
  {\doibase 10.1016/j.nuclphysbps.2004.11.213} {\bibfield  {journal} {\bibinfo
  {journal} {Nucl. Phys. Proc. Suppl.}\ }\textbf {\bibinfo {volume} {139}},\
  \bibinfo {pages} {158} (\bibinfo {year} {2005}{\natexlab{b}})},\ \bibinfo
  {note} {[,158(2004)]},\ \Eprint {http://arxiv.org/abs/hep-ph/0408106}
  {arXiv:hep-ph/0408106 [hep-ph]} \BibitemShut {NoStop}%
\bibitem [{\citenamefont {Valverde}\ \emph {et~al.}(2006)\citenamefont
  {Valverde}, \citenamefont {Amaro}, \citenamefont {Nieves},\ and\
  \citenamefont {Maieron}}]{Valverde:2006yi}%
  \BibitemOpen
  \bibfield  {author} {\bibinfo {author} {\bibfnamefont {M.}~\bibnamefont
  {Valverde}}, \bibinfo {author} {\bibfnamefont {J.~E.}\ \bibnamefont {Amaro}},
  \bibinfo {author} {\bibfnamefont {J.}~\bibnamefont {Nieves}}, \ and\ \bibinfo
  {author} {\bibfnamefont {C.}~\bibnamefont {Maieron}},\ }\href {\doibase
  10.1016/j.physletb.2006.08.087} {\bibfield  {journal} {\bibinfo  {journal}
  {Phys. Lett. B}\ }\textbf {\bibinfo {volume} {642}},\ \bibinfo {pages} {218}
  (\bibinfo {year} {2006})},\ \Eprint {http://arxiv.org/abs/nucl-th/0606042}
  {arXiv:nucl-th/0606042} \BibitemShut {NoStop}%
\bibitem [{\citenamefont {Sobczyk}\ \emph {et~al.}(2019)\citenamefont
  {Sobczyk}, \citenamefont {Rocco},\ and\ \citenamefont
  {Nieves}}]{Sobczyk:2019urm}%
  \BibitemOpen
  \bibfield  {author} {\bibinfo {author} {\bibfnamefont {J.~E.}\ \bibnamefont
  {Sobczyk}}, \bibinfo {author} {\bibfnamefont {N.}~\bibnamefont {Rocco}}, \
  and\ \bibinfo {author} {\bibfnamefont {J.}~\bibnamefont {Nieves}},\ }\href
  {\doibase 10.1103/PhysRevC.100.035501} {\bibfield  {journal} {\bibinfo
  {journal} {Phys. Rev. C}\ }\textbf {\bibinfo {volume} {100}},\ \bibinfo
  {pages} {035501} (\bibinfo {year} {2019})},\ \Eprint
  {http://arxiv.org/abs/1906.05656} {arXiv:1906.05656 [nucl-th]} \BibitemShut
  {NoStop}%
\bibitem [{\citenamefont {Fatima}\ \emph {et~al.}(2020)\citenamefont {Fatima},
  \citenamefont {Sajjad~Athar},\ and\ \citenamefont {Singh}}]{Fatima:2020pvv}%
  \BibitemOpen
  \bibfield  {author} {\bibinfo {author} {\bibfnamefont {A.}~\bibnamefont
  {Fatima}}, \bibinfo {author} {\bibfnamefont {M.}~\bibnamefont
  {Sajjad~Athar}}, \ and\ \bibinfo {author} {\bibfnamefont {S.~K.}\
  \bibnamefont {Singh}},\ }\href {\doibase 10.1103/PhysRevD.102.113009}
  {\bibfield  {journal} {\bibinfo  {journal} {Phys. Rev. D}\ }\textbf {\bibinfo
  {volume} {102}},\ \bibinfo {pages} {113009} (\bibinfo {year} {2020})},\
  \Eprint {http://arxiv.org/abs/2010.10311} {arXiv:2010.10311 [hep-ph]}
  \BibitemShut {NoStop}%
\bibitem [{\citenamefont {Hern\'andez}\ \emph {et~al.}(2022)\citenamefont
  {Hern\'andez}, \citenamefont {Nieves}, \citenamefont {S\'anchez},\ and\
  \citenamefont {Sobczyk}}]{Hernandez:2022nmp}%
  \BibitemOpen
  \bibfield  {author} {\bibinfo {author} {\bibfnamefont {E.}~\bibnamefont
  {Hern\'andez}}, \bibinfo {author} {\bibfnamefont {J.}~\bibnamefont {Nieves}},
  \bibinfo {author} {\bibfnamefont {F.}~\bibnamefont {S\'anchez}}, \ and\
  \bibinfo {author} {\bibfnamefont {J.~E.}\ \bibnamefont {Sobczyk}},\ }\href
  {\doibase 10.1016/j.physletb.2022.137046} {\bibfield  {journal} {\bibinfo
  {journal} {Phys. Lett. B}\ }\textbf {\bibinfo {volume} {829}},\ \bibinfo
  {pages} {137046} (\bibinfo {year} {2022})},\ \Eprint
  {http://arxiv.org/abs/2202.07539} {arXiv:2202.07539 [hep-ph]} \BibitemShut
  {NoStop}%
\bibitem [{\citenamefont {Isaacson}\ \emph {et~al.}(2023)\citenamefont
  {Isaacson}, \citenamefont {H\"oche}, \citenamefont {Siegert},\ and\
  \citenamefont {Wang}}]{Isaacson:2023gwp}%
  \BibitemOpen
  \bibfield  {author} {\bibinfo {author} {\bibfnamefont {J.}~\bibnamefont
  {Isaacson}}, \bibinfo {author} {\bibfnamefont {S.}~\bibnamefont {H\"oche}},
  \bibinfo {author} {\bibfnamefont {F.}~\bibnamefont {Siegert}}, \ and\
  \bibinfo {author} {\bibfnamefont {S.}~\bibnamefont {Wang}},\ }\href@noop {}
  {\  (\bibinfo {year} {2023})},\ \Eprint {http://arxiv.org/abs/2303.08104}
  {arXiv:2303.08104 [hep-ph]} \BibitemShut {NoStop}%
\bibitem [{\citenamefont {Li}\ \emph {et~al.}(2018)\citenamefont {Li} \emph
  {et~al.}}]{Li:2017dbe}%
  \BibitemOpen
  \bibfield  {author} {\bibinfo {author} {\bibfnamefont {Z.}~\bibnamefont {Li}}
  \emph {et~al.} (\bibinfo {collaboration} {Super-Kamiokande}),\ }\href
  {\doibase 10.1103/PhysRevD.98.052006} {\bibfield  {journal} {\bibinfo
  {journal} {Phys. Rev. D}\ }\textbf {\bibinfo {volume} {98}},\ \bibinfo
  {pages} {052006} (\bibinfo {year} {2018})},\ \Eprint
  {http://arxiv.org/abs/1711.09436} {arXiv:1711.09436 [hep-ex]} \BibitemShut
  {NoStop}%
\bibitem [{\citenamefont {Akbar}\ \emph {et~al.}(2016)\citenamefont {Akbar},
  \citenamefont {Rafi~Alam}, \citenamefont {Sajjad~Athar},\ and\ \citenamefont
  {Singh}}]{Akbar:2016awk}%
  \BibitemOpen
  \bibfield  {author} {\bibinfo {author} {\bibfnamefont {F.}~\bibnamefont
  {Akbar}}, \bibinfo {author} {\bibfnamefont {M.}~\bibnamefont {Rafi~Alam}},
  \bibinfo {author} {\bibfnamefont {M.}~\bibnamefont {Sajjad~Athar}}, \ and\
  \bibinfo {author} {\bibfnamefont {S.~K.}\ \bibnamefont {Singh}},\ }\href
  {\doibase 10.1103/PhysRevD.94.114031} {\bibfield  {journal} {\bibinfo
  {journal} {Phys. Rev. D}\ }\textbf {\bibinfo {volume} {94}},\ \bibinfo
  {pages} {114031} (\bibinfo {year} {2016})},\ \Eprint
  {http://arxiv.org/abs/1608.02103} {arXiv:1608.02103 [hep-ph]} \BibitemShut
  {NoStop}%
\bibitem [{\citenamefont {Fatima}\ \emph
  {et~al.}(2018{\natexlab{a}})\citenamefont {Fatima}, \citenamefont
  {Sajjad~Athar},\ and\ \citenamefont {Singh}}]{Fatima:2018tzs}%
  \BibitemOpen
  \bibfield  {author} {\bibinfo {author} {\bibfnamefont {A.}~\bibnamefont
  {Fatima}}, \bibinfo {author} {\bibfnamefont {M.}~\bibnamefont
  {Sajjad~Athar}}, \ and\ \bibinfo {author} {\bibfnamefont {S.~K.}\
  \bibnamefont {Singh}},\ }\href {\doibase 10.1103/PhysRevD.98.033005}
  {\bibfield  {journal} {\bibinfo  {journal} {Phys. Rev. D}\ }\textbf {\bibinfo
  {volume} {98}},\ \bibinfo {pages} {033005} (\bibinfo {year}
  {2018}{\natexlab{a}})},\ \Eprint {http://arxiv.org/abs/1806.08597}
  {arXiv:1806.08597 [hep-ph]} \BibitemShut {NoStop}%
\bibitem [{\citenamefont {Graczyk}\ and\ \citenamefont
  {Kowal}(2021)}]{Graczyk:2021oyl}%
  \BibitemOpen
  \bibfield  {author} {\bibinfo {author} {\bibfnamefont {K.~M.}\ \bibnamefont
  {Graczyk}}\ and\ \bibinfo {author} {\bibfnamefont {B.~E.}\ \bibnamefont
  {Kowal}},\ }\href {\doibase 10.1103/PhysRevD.104.033005} {\bibfield
  {journal} {\bibinfo  {journal} {Phys. Rev. D}\ }\textbf {\bibinfo {volume}
  {104}},\ \bibinfo {pages} {033005} (\bibinfo {year} {2021})},\ \Eprint
  {http://arxiv.org/abs/2106.11383} {arXiv:2106.11383 [hep-ph]} \BibitemShut
  {NoStop}%
\bibitem [{\citenamefont {Graczyk}\ and\ \citenamefont
  {Kowal}(2023)}]{Graczyk:2023lrm}%
  \BibitemOpen
  \bibfield  {author} {\bibinfo {author} {\bibfnamefont {K.~M.}\ \bibnamefont
  {Graczyk}}\ and\ \bibinfo {author} {\bibfnamefont {B.~E.}\ \bibnamefont
  {Kowal}},\ }\href@noop {} {\  (\bibinfo {year} {2023})},\ \Eprint
  {http://arxiv.org/abs/2307.00661} {arXiv:2307.00661 [hep-ph]} \BibitemShut
  {NoStop}%
\bibitem [{\citenamefont {Bilenky}\ and\ \citenamefont
  {Christova}(2013{\natexlab{a}})}]{Bilenky:2013fra}%
  \BibitemOpen
  \bibfield  {author} {\bibinfo {author} {\bibfnamefont {S.~M.}\ \bibnamefont
  {Bilenky}}\ and\ \bibinfo {author} {\bibfnamefont {E.}~\bibnamefont
  {Christova}},\ }\href {\doibase 10.1088/0954-3899/40/7/075004} {\bibfield
  {journal} {\bibinfo  {journal} {J. Phys. G}\ }\textbf {\bibinfo {volume}
  {40}},\ \bibinfo {pages} {075004} (\bibinfo {year} {2013}{\natexlab{a}})},\
  \Eprint {http://arxiv.org/abs/1303.3710} {arXiv:1303.3710 [hep-ph]}
  \BibitemShut {NoStop}%
\bibitem [{\citenamefont {Bilenky}\ and\ \citenamefont
  {Christova}(2013{\natexlab{b}})}]{Bilenky:2013iua}%
  \BibitemOpen
  \bibfield  {author} {\bibinfo {author} {\bibfnamefont {S.~M.}\ \bibnamefont
  {Bilenky}}\ and\ \bibinfo {author} {\bibfnamefont {E.}~\bibnamefont
  {Christova}},\ }\href {\doibase 10.1134/S154747711307011X} {\bibfield
  {journal} {\bibinfo  {journal} {Phys. Part. Nucl. Lett.}\ }\textbf {\bibinfo
  {volume} {10}},\ \bibinfo {pages} {651} (\bibinfo {year}
  {2013}{\natexlab{b}})},\ \Eprint {http://arxiv.org/abs/1307.7275}
  {arXiv:1307.7275 [hep-ph]} \BibitemShut {NoStop}%
\bibitem [{\citenamefont {Miranda}\ and\ \citenamefont
  {Nunokawa}(2015)}]{Miranda:2015dra}%
  \BibitemOpen
  \bibfield  {author} {\bibinfo {author} {\bibfnamefont {O.~G.}\ \bibnamefont
  {Miranda}}\ and\ \bibinfo {author} {\bibfnamefont {H.}~\bibnamefont
  {Nunokawa}},\ }\href {\doibase 10.1088/1367-2630/17/9/095002} {\bibfield
  {journal} {\bibinfo  {journal} {New J. Phys.}\ }\textbf {\bibinfo {volume}
  {17}},\ \bibinfo {pages} {095002} (\bibinfo {year} {2015})},\ \Eprint
  {http://arxiv.org/abs/1505.06254} {arXiv:1505.06254 [hep-ph]} \BibitemShut
  {NoStop}%
\bibitem [{\citenamefont {Tomalak}\ \emph
  {et~al.}(2022{\natexlab{a}})\citenamefont {Tomalak}, \citenamefont {Chen},
  \citenamefont {Hill},\ and\ \citenamefont {McFarland}}]{Tomalak:2021hec}%
  \BibitemOpen
  \bibfield  {author} {\bibinfo {author} {\bibfnamefont {O.}~\bibnamefont
  {Tomalak}}, \bibinfo {author} {\bibfnamefont {Q.}~\bibnamefont {Chen}},
  \bibinfo {author} {\bibfnamefont {R.~J.}\ \bibnamefont {Hill}}, \ and\
  \bibinfo {author} {\bibfnamefont {K.~S.}\ \bibnamefont {McFarland}},\ }\href
  {\doibase 10.1038/s41467-022-32974-x} {\bibfield  {journal} {\bibinfo
  {journal} {Nature Commun.}\ }\textbf {\bibinfo {volume} {13}},\ \bibinfo
  {pages} {5286} (\bibinfo {year} {2022}{\natexlab{a}})},\ \Eprint
  {http://arxiv.org/abs/2105.07939} {arXiv:2105.07939 [hep-ph]} \BibitemShut
  {NoStop}%
\bibitem [{\citenamefont {Tomalak}\ \emph
  {et~al.}(2022{\natexlab{b}})\citenamefont {Tomalak}, \citenamefont {Chen},
  \citenamefont {Hill}, \citenamefont {McFarland},\ and\ \citenamefont
  {Wret}}]{Tomalak:2022xup}%
  \BibitemOpen
  \bibfield  {author} {\bibinfo {author} {\bibfnamefont {O.}~\bibnamefont
  {Tomalak}}, \bibinfo {author} {\bibfnamefont {Q.}~\bibnamefont {Chen}},
  \bibinfo {author} {\bibfnamefont {R.~J.}\ \bibnamefont {Hill}}, \bibinfo
  {author} {\bibfnamefont {K.~S.}\ \bibnamefont {McFarland}}, \ and\ \bibinfo
  {author} {\bibfnamefont {C.}~\bibnamefont {Wret}},\ }\href {\doibase
  10.1103/PhysRevD.106.093006} {\bibfield  {journal} {\bibinfo  {journal}
  {Phys. Rev. D}\ }\textbf {\bibinfo {volume} {106}},\ \bibinfo {pages}
  {093006} (\bibinfo {year} {2022}{\natexlab{b}})},\ \Eprint
  {http://arxiv.org/abs/2204.11379} {arXiv:2204.11379 [hep-ph]} \BibitemShut
  {NoStop}%
\bibitem [{\citenamefont {Abbaslu}\ \emph {et~al.}(2024)\citenamefont
  {Abbaslu}, \citenamefont {Dehpour}, \citenamefont {Farzan},\ and\
  \citenamefont {Safari}}]{Abbaslu:2024jzo}%
  \BibitemOpen
  \bibfield  {author} {\bibinfo {author} {\bibfnamefont {S.}~\bibnamefont
  {Abbaslu}}, \bibinfo {author} {\bibfnamefont {M.}~\bibnamefont {Dehpour}},
  \bibinfo {author} {\bibfnamefont {Y.}~\bibnamefont {Farzan}}, \ and\ \bibinfo
  {author} {\bibfnamefont {S.}~\bibnamefont {Safari}},\ }\href@noop {} {\
  (\bibinfo {year} {2024})},\ \Eprint {http://arxiv.org/abs/2412.13349}
  {arXiv:2412.13349 [hep-ph]} \BibitemShut {NoStop}%
\bibitem [{\citenamefont {Maas}\ and\ \citenamefont
  {Paschke}(2017)}]{Maas:2017snj}%
  \BibitemOpen
  \bibfield  {author} {\bibinfo {author} {\bibfnamefont {F.~E.}\ \bibnamefont
  {Maas}}\ and\ \bibinfo {author} {\bibfnamefont {K.~D.}\ \bibnamefont
  {Paschke}},\ }\href {\doibase 10.1016/j.ppnp.2016.11.001} {\bibfield
  {journal} {\bibinfo  {journal} {Prog. Part. Nucl. Phys.}\ }\textbf {\bibinfo
  {volume} {95}},\ \bibinfo {pages} {209} (\bibinfo {year} {2017})}\BibitemShut
  {NoStop}%
\bibitem [{\citenamefont {Garvey}\ \emph {et~al.}(1993)\citenamefont {Garvey},
  \citenamefont {Louis},\ and\ \citenamefont {White}}]{Garvey:1992cg}%
  \BibitemOpen
  \bibfield  {author} {\bibinfo {author} {\bibfnamefont {G.~T.}\ \bibnamefont
  {Garvey}}, \bibinfo {author} {\bibfnamefont {W.~C.}\ \bibnamefont {Louis}}, \
  and\ \bibinfo {author} {\bibfnamefont {D.~H.}\ \bibnamefont {White}},\ }\href
  {\doibase 10.1103/PhysRevC.48.761} {\bibfield  {journal} {\bibinfo  {journal}
  {Phys. Rev. C}\ }\textbf {\bibinfo {volume} {48}},\ \bibinfo {pages} {761}
  (\bibinfo {year} {1993})}\BibitemShut {NoStop}%
\bibitem [{\citenamefont {Aoki}\ \emph {et~al.}(2022)\citenamefont {Aoki} \emph
  {et~al.}}]{flavourLatticeAveragingGroupFLAG:2021npn}%
  \BibitemOpen
  \bibfield  {author} {\bibinfo {author} {\bibfnamefont {Y.}~\bibnamefont
  {Aoki}} \emph {et~al.} (\bibinfo {collaboration} {Flavour Lattice Averaging
  Group (FLAG)}),\ }\href {\doibase 10.1140/epjc/s10052-022-10536-1} {\bibfield
   {journal} {\bibinfo  {journal} {Eur. Phys. J. C}\ }\textbf {\bibinfo
  {volume} {82}},\ \bibinfo {pages} {869} (\bibinfo {year} {2022})},\ \Eprint
  {http://arxiv.org/abs/2111.09849} {arXiv:2111.09849 [hep-lat]} \BibitemShut
  {NoStop}%
\bibitem [{\citenamefont {Alexandrou}\ \emph {et~al.}(2020)\citenamefont
  {Alexandrou}, \citenamefont {Bacchio}, \citenamefont {Constantinou},
  \citenamefont {Finkenrath}, \citenamefont {Hadjiyiannakou}, \citenamefont
  {Jansen},\ and\ \citenamefont {Koutsou}}]{Alexandrou:2019olr}%
  \BibitemOpen
  \bibfield  {author} {\bibinfo {author} {\bibfnamefont {C.}~\bibnamefont
  {Alexandrou}}, \bibinfo {author} {\bibfnamefont {S.}~\bibnamefont {Bacchio}},
  \bibinfo {author} {\bibfnamefont {M.}~\bibnamefont {Constantinou}}, \bibinfo
  {author} {\bibfnamefont {J.}~\bibnamefont {Finkenrath}}, \bibinfo {author}
  {\bibfnamefont {K.}~\bibnamefont {Hadjiyiannakou}}, \bibinfo {author}
  {\bibfnamefont {K.}~\bibnamefont {Jansen}}, \ and\ \bibinfo {author}
  {\bibfnamefont {G.}~\bibnamefont {Koutsou}},\ }\href {\doibase
  10.1103/PhysRevD.101.031501} {\bibfield  {journal} {\bibinfo  {journal}
  {Phys. Rev. D}\ }\textbf {\bibinfo {volume} {101}},\ \bibinfo {pages}
  {031501} (\bibinfo {year} {2020})},\ \Eprint
  {http://arxiv.org/abs/1909.10744} {arXiv:1909.10744 [hep-lat]} \BibitemShut
  {NoStop}%
\bibitem [{\citenamefont {Alexandrou}\ \emph {et~al.}(2021)\citenamefont
  {Alexandrou}, \citenamefont {Bacchio}, \citenamefont {Constantinou},
  \citenamefont {Hadjiyiannakou}, \citenamefont {Jansen},\ and\ \citenamefont
  {Koutsou}}]{Alexandrou:2021wzv}%
  \BibitemOpen
  \bibfield  {author} {\bibinfo {author} {\bibfnamefont {C.}~\bibnamefont
  {Alexandrou}}, \bibinfo {author} {\bibfnamefont {S.}~\bibnamefont {Bacchio}},
  \bibinfo {author} {\bibfnamefont {M.}~\bibnamefont {Constantinou}}, \bibinfo
  {author} {\bibfnamefont {K.}~\bibnamefont {Hadjiyiannakou}}, \bibinfo
  {author} {\bibfnamefont {K.}~\bibnamefont {Jansen}}, \ and\ \bibinfo {author}
  {\bibfnamefont {G.}~\bibnamefont {Koutsou}},\ }\href {\doibase
  10.1103/PhysRevD.104.074503} {\bibfield  {journal} {\bibinfo  {journal}
  {Phys. Rev. D}\ }\textbf {\bibinfo {volume} {104}},\ \bibinfo {pages}
  {074503} (\bibinfo {year} {2021})},\ \Eprint
  {http://arxiv.org/abs/2106.13468} {arXiv:2106.13468 [hep-lat]} \BibitemShut
  {NoStop}%
\bibitem [{\citenamefont {Cabibbo}(1964)}]{Cabibbo:1964zza}%
  \BibitemOpen
  \bibfield  {author} {\bibinfo {author} {\bibfnamefont {N.}~\bibnamefont
  {Cabibbo}},\ }\href {\doibase 10.1016/0031-9163(64)91138-2} {\bibfield
  {journal} {\bibinfo  {journal} {Phys. Lett.}\ }\textbf {\bibinfo {volume}
  {12}},\ \bibinfo {pages} {137} (\bibinfo {year} {1964})}\BibitemShut
  {NoStop}%
\bibitem [{\citenamefont {Bilenky}(1995)}]{Bilenky:1995zq}%
  \BibitemOpen
  \bibfield  {author} {\bibinfo {author} {\bibfnamefont {S.~M.}\ \bibnamefont
  {Bilenky}},\ }\href@noop {} {\emph {\bibinfo {title} {{Introduction to
  Feynman diagrams and electroweak interactions physics}}}}\ (\bibinfo {year}
  {1995})\BibitemShut {NoStop}%
\bibitem [{\citenamefont {Fatima}\ \emph
  {et~al.}(2018{\natexlab{b}})\citenamefont {Fatima}, \citenamefont
  {Sajjad~Athar},\ and\ \citenamefont {Singh}}]{Fatima:2018gjy}%
  \BibitemOpen
  \bibfield  {author} {\bibinfo {author} {\bibfnamefont {A.}~\bibnamefont
  {Fatima}}, \bibinfo {author} {\bibfnamefont {M.}~\bibnamefont
  {Sajjad~Athar}}, \ and\ \bibinfo {author} {\bibfnamefont {S.~K.}\
  \bibnamefont {Singh}},\ }\href {\doibase 10.1140/epja/i2018-12534-2}
  {\bibfield  {journal} {\bibinfo  {journal} {Eur. Phys. J. A}\ }\textbf
  {\bibinfo {volume} {54}},\ \bibinfo {pages} {95} (\bibinfo {year}
  {2018}{\natexlab{b}})},\ \Eprint {http://arxiv.org/abs/1802.04469}
  {arXiv:1802.04469 [hep-ph]} \BibitemShut {NoStop}%
\bibitem [{\citenamefont {Sajjad~Athar}\ \emph {et~al.}(2023)\citenamefont
  {Sajjad~Athar}, \citenamefont {Fatima},\ and\ \citenamefont
  {Singh}}]{SajjadAthar:2022pjt}%
  \BibitemOpen
  \bibfield  {author} {\bibinfo {author} {\bibfnamefont {M.}~\bibnamefont
  {Sajjad~Athar}}, \bibinfo {author} {\bibfnamefont {A.}~\bibnamefont
  {Fatima}}, \ and\ \bibinfo {author} {\bibfnamefont {S.~K.}\ \bibnamefont
  {Singh}},\ }\href {\doibase 10.1016/j.ppnp.2022.104019} {\bibfield  {journal}
  {\bibinfo  {journal} {Prog. Part. Nucl. Phys.}\ }\textbf {\bibinfo {volume}
  {129}},\ \bibinfo {pages} {104019} (\bibinfo {year} {2023})},\ \Eprint
  {http://arxiv.org/abs/2206.13792} {arXiv:2206.13792 [hep-ph]} \BibitemShut
  {NoStop}%
\bibitem [{\citenamefont {Singh}\ and\ \citenamefont
  {Arenhovel}(1986)}]{Singh:1986xh}%
  \BibitemOpen
  \bibfield  {author} {\bibinfo {author} {\bibfnamefont {S.~K.}\ \bibnamefont
  {Singh}}\ and\ \bibinfo {author} {\bibfnamefont {H.}~\bibnamefont
  {Arenhovel}},\ }\href {\doibase 10.1007/BF01294589} {\bibfield  {journal}
  {\bibinfo  {journal} {Z. Phys. A}\ }\textbf {\bibinfo {volume} {324}},\
  \bibinfo {pages} {347} (\bibinfo {year} {1986})}\BibitemShut {NoStop}%
\bibitem [{\citenamefont {Shen}\ \emph {et~al.}(2012)\citenamefont {Shen},
  \citenamefont {Marcucci}, \citenamefont {Carlson}, \citenamefont {Gandolfi},\
  and\ \citenamefont {Schiavilla}}]{Shen:2012xz}%
  \BibitemOpen
  \bibfield  {author} {\bibinfo {author} {\bibfnamefont {G.}~\bibnamefont
  {Shen}}, \bibinfo {author} {\bibfnamefont {L.~E.}\ \bibnamefont {Marcucci}},
  \bibinfo {author} {\bibfnamefont {J.}~\bibnamefont {Carlson}}, \bibinfo
  {author} {\bibfnamefont {S.}~\bibnamefont {Gandolfi}}, \ and\ \bibinfo
  {author} {\bibfnamefont {R.}~\bibnamefont {Schiavilla}},\ }\href {\doibase
  10.1103/PhysRevC.86.035503} {\bibfield  {journal} {\bibinfo  {journal} {Phys.
  Rev. C}\ }\textbf {\bibinfo {volume} {86}},\ \bibinfo {pages} {035503}
  (\bibinfo {year} {2012})},\ \Eprint {http://arxiv.org/abs/1205.4337}
  {arXiv:1205.4337 [nucl-th]} \BibitemShut {NoStop}%
\bibitem [{\citenamefont {Alvarez-Ruso}\ \emph {et~al.}(2014)\citenamefont
  {Alvarez-Ruso}, \citenamefont {Hayato},\ and\ \citenamefont
  {Nieves}}]{Alvarez-Ruso:2014bla}%
  \BibitemOpen
  \bibfield  {author} {\bibinfo {author} {\bibfnamefont {L.}~\bibnamefont
  {Alvarez-Ruso}}, \bibinfo {author} {\bibfnamefont {Y.}~\bibnamefont
  {Hayato}}, \ and\ \bibinfo {author} {\bibfnamefont {J.}~\bibnamefont
  {Nieves}},\ }\href {\doibase 10.1088/1367-2630/16/7/075015} {\bibfield
  {journal} {\bibinfo  {journal} {New J. Phys.}\ }\textbf {\bibinfo {volume}
  {16}},\ \bibinfo {pages} {075015} (\bibinfo {year} {2014})},\ \Eprint
  {http://arxiv.org/abs/1403.2673} {arXiv:1403.2673 [hep-ph]} \BibitemShut
  {NoStop}%
\bibitem [{\citenamefont {Galster}\ \emph {et~al.}(1971)\citenamefont
  {Galster}, \citenamefont {Klein}, \citenamefont {Moritz}, \citenamefont
  {Schmidt}, \citenamefont {Wegener},\ and\ \citenamefont
  {Bleckwenn}}]{Galster:1971kv}%
  \BibitemOpen
  \bibfield  {author} {\bibinfo {author} {\bibfnamefont {S.}~\bibnamefont
  {Galster}}, \bibinfo {author} {\bibfnamefont {H.}~\bibnamefont {Klein}},
  \bibinfo {author} {\bibfnamefont {J.}~\bibnamefont {Moritz}}, \bibinfo
  {author} {\bibfnamefont {K.~H.}\ \bibnamefont {Schmidt}}, \bibinfo {author}
  {\bibfnamefont {D.}~\bibnamefont {Wegener}}, \ and\ \bibinfo {author}
  {\bibfnamefont {J.}~\bibnamefont {Bleckwenn}},\ }\href {\doibase
  10.1016/0550-3213(71)90068-X} {\bibfield  {journal} {\bibinfo  {journal}
  {Nucl. Phys. B}\ }\textbf {\bibinfo {volume} {32}},\ \bibinfo {pages} {221}
  (\bibinfo {year} {1971})}\BibitemShut {NoStop}%
\end{thebibliography}%
\bibliographystyle{apsrev4-1}

\end{document}